\documentclass[%
reprint,
superscriptaddress,
nofootinbib,
amsmath,
amssymb,
aps,
floatfix,
]{revtex4-1}
\pdfoutput=1
\usepackage{graphicx}
\usepackage{dcolumn}
\usepackage{bm}
\usepackage{color}
\usepackage{adjustbox}
\usepackage{mathtools}
\usepackage{lipsum}
\usepackage{endnotes}
\usepackage{natbib}
\usepackage[english]{babel}
\usepackage{hyperref}
\usepackage[utf8]{inputenc}
\usepackage{csquotes}

\begin{document}

\title{Structural origins of electronic conduction in amorphous copper-doped alumina}

 \author{K. N. Subedi}
 \email{ks173214@ohio.edu}
 \affiliation{Department of Physics and Astronomy, Ohio University, Athens, OH 45701, USA}

 \author{K. Prasai}
 \affiliation{E. L. Ginzton Lab, Stanford University, Stanford, CA 94305, USA}
 
 \author{M. N. Kozicki}
 \affiliation{School of Electrical, Computer and Energy Engineering, Arizona State University, Tempe, AZ 85287, USA}
 
 \author{D. A. Drabold}
 \email{drabold@ohio.edu}
 \affiliation{Department of Physics and Astronomy, Ohio University, Athens, OH 45701, USA}
 
\date{\today}

\begin{abstract}
We perform an {\it ab initio} modeling of amorphous copper-doped alumina (a-Al$_2$O$_3$:Cu), a prospective memory material based on resistance switching, and study the structural origin of electronic conduction in this material. We generate molecular dynamics based models of a-Al$_2$O$_3$:Cu at various Cu-concentrations and study the structural, electronic and vibrational properties as a function of Cu-concentration. Cu atoms show a strong tendency to cluster in the alumina host, and metallize the system by filling the band gap uniformly for higher Cu-concentrations. We also study thermal fluctuations of the HOMO-LUMO energy splitting and observe the time evolution of the size of the band gap, which can be expected to have an important impact on the conductivity. We perform a numerical computation of conduction pathways, and show its explicit dependence on Cu connectivity in the host. We present an analysis of ion dynamics and structural aspects of localization of classical normal modes in our models. 
\end{abstract}

\maketitle

\section{Introduction}
Non-volatile memory devices based on resistive switching characteristics have been studied since the late 1960s \cite{history1}. In these devices, application of an external bias potential across an electrolyte changes the electrical conductivity of the electrolyte by changing its structure. This process is reversible and can be performed in the time scale of nanoseconds. Three types of resistive random access memory (RRAM) devices have been studied in detail \cite{history2} and these include RRAM based on oxygen vacancies, RRAM based on thermo-chemical effects and RRAM based on the electro-chemical metallization. The later class of devices are also called conducting bridge random access memory or CBRAM. The CBRAM devices are composed of a thin solid electrolyte layer placed between an oxidizable anode ($eg.$ Cu, Ag or TiN) and an inert cathode ($eg.$ W or Pt). The Cu, in its ionic state, is converted into the conducting \enquote{filament} by the applied field: the ions are reduced by electrons flowing from the cathode to leave them in their metallic form, although other counter ions (e.g., OH-) may also be involved in this process \cite{Valov_2013}. With the application of a reverse bias, the connectivity of the cluster can be destroyed, and the device is put into a highly electronically resistive state. The details of the mechanism of CBRAMs have been described elsewhere \cite{kozicki_review,cbram_stefan}. The performance of CBRAM devices has been studied with several materials as the solid electrolyte which include chalcogenides \cite{chalco1,chalco2_gese2}, insulating metal oxides \cite{chen_sio2,tsuruoka,ta2o5_oxide,xu_alumina,al2o31,al2o32,sumeet} and bilayer materials \cite{bilayer_tsai,bilayer_barci}. CBRAM devices have demonstrated excellent performance in terms of operational voltage, read/write speed, endurance and data retention. Among the host materials reviewed for CBRAM devices, alumina (Al$_2$O$_3$) shows particular promise. It has a high dielectric constant, large band gap, and its amorphous phase is highly stable \cite{kittl_high_k,eklund_thermalstability_al2o3}. The experimental results for CBRAM devices based on Cu alloyed with Al$_2$O$_3$ have shown that the cell exhibits highly controlled set and reset operations, fast pulse programming (10 ns) at low voltage (\textless3 V) and low-current (10 $\mu$A) with 10$^6$ cycles per second for the writing speed \cite{al2o31}.\\
\\
In this paper, we use \textit{ab initio} molecular dynamics (AIMD) to generate atomic models of a-Al$_2$O$_3$:Cu and investigate the microscopic origin of elctronic conduction in this material. The work presented in this paper shows that an increase in local Cu-concentration can result in stable conducting pathways due to the strong tendency of Cu atoms to cluster in the ionic host. This would lead to a highly stable low resistance state (LRS) for high copper concentration, which does indeed seem to be the case for copper-alumina devices \cite{al2o31}. We study the electronic properties for these models and are able to crudely estimate the local concentration of Cu above which CBRAM device switch to the LRS. We present the numerical computation of conduction-active parts of the network based on our recent work on computing space projected conductivity (SPC) \cite{rrl_prasai18}, and show that the strong electron-lattice coupling for electron states near the gap leads to interesting and substantial thermally induced conductivity fluctuations on a picosecond time scale.\\
\\
The rest of the paper is organized as follows. Section II describes computational details used to create the structures and also the details of our method to obtain the SPC. Section III includes results where we discuss structural, electronic and vibrational properties of the models in different subsections. Section IV provides the conclusions.

\section{\label{sec:level2}Computations}
\subsection{\label{sec:level2_a}Model Generation}
 In this work, we use AIMD to generate four atomic models with the composition of (a-Al$_2$O$_3$)$_{1-n}$Cu$_{n}$ with $n=$ 0, 0.1, 0.2 and 0.3. We used a density of 3.175 g/cm$^3$ for a-Al$_2$O$_3$ \cite{gutierrez,vasistha_al2o3}. For the Cu-doped models, we referred to the literature \cite{cudoped_reference} to make an initial guess, then carried out a zero-pressure relaxation to correct/optimize the result. For each model, we began by taking a cubic supercell of 200 atoms with randomly initialized positions of the atoms.
 ~Plane wave density functional calculations were performed using the VASP package \cite{vasp} and projector-augmented wave (PAW) \cite{PAW1,PAW2} potentials within the local density approximation (LDA) \cite{LDA} using periodic boundary conditions. We used a kinetic energy cutoff of 420 eV and the $\Gamma$-point to sample the Brillouin zone. A time step of 1.5 fs was used and the temperature was controlled by a Nos\'e-Hoover thermostat throughout.
 
 We performed a melt-quench simulation \cite{dadepj} with a starting temperature of 3500 K. After annealing the \enquote{hot liquid} for 7.5 ps at 3500 K, we cooled each model to 2600 K at a rate of 0.27 K/fs as discussed in reference \cite{cooling_rate_al2o3} and then equilibrated for 10 ps. Each model was then quenched to 300 K at the same cooling rate 0.27 K/fs and further equilibrated for another 10 ps. Zero pressure relaxations were used to determine the final densities for Cu-doped models. The final force between the atoms is no more than 0.01 eV/atom. The initial and final densities are provided in table \ref{table1}.
 
\begin{table}[h]
\caption{Initial and final densities of a-Al$_2$O$_3$:Cu models}
\begin{center}
\begin{tabular*}{0.9\linewidth}{@{\extracolsep\fill}cccc@{\extracolsep\fill}}
\hline
\textbf{Cu content}&\textbf{Mol. Formula} &\textbf{$\rho_{in}$(g/cc)}&\textbf{$\rho_{f}$(g/cc)}\\
\hline
0\% & (Al$_2$O$_{3}$)$_{1.00}$Cu$_{0.00}$ & 3.175 & 3.175  \\
10\% & (Al$_2$O$_{3}$)$_{0.90}$Cu$_{0.10}$ & 3.58 & 3.75  \\
20\% & (Al$_2$O$_{3}$)$_{0.80}$Cu$_{0.20}$ & 3.78 & 3.99  \\
30\% & (Al$_2$O$_{3}$)$_{0.70}$Cu$_{0.30}$ & 4.53 & 4.82  \\
\hline
\end{tabular*}
\end{center}
\label{table1}
\end{table}

\subsection{\label{sec:level2_b}Spatial Projection of Electronic Conductivity}
In this section, we discuss a method to obtain a space projected electronic conductivity. We discuss the method in detail in Ref. \cite{rrl_prasai18}. We begin by writing the diagonal elements of the conductivity tensor for each k-point \textbf{k} and frequency $\omega$ using the standard Kubo-Greenwood formula KGF \cite{kubo,greenwood} as:
\begin{equation}
\begin{aligned}
\sigma_{\bf k}(\omega) =\frac{2\pi e^2}{3m^2\omega \Omega}\sum_{i,j} \sum_\alpha [f(\epsilon_{i,\bf k})-f(\epsilon_{j,\bf k})]\\
{\mid \langle \psi_{j, \bf k}|p^\alpha|\psi_{i,\bf k} \rangle \mid}^2 \delta(\epsilon_{j,{ \bf k}}-\epsilon_{i,{\bf k}}-\hbar \omega)
\label{kgf_eqn}
\end{aligned}
\end{equation}
In the above equation (\ref{kgf_eqn}), \textit{e} and \textit{m} represent the charge and mass of the electron respectively. $\Omega$ represents the volume of the supercell. We average over diagonal elements of conductivity tensor($\alpha = x,y,z$). $\psi_{i,\bf k}$ is the Kohn-Sham orbital associated with energy $\epsilon_{i,\bf k}$ and $f(\epsilon_{i,\bf k})$ denotes the Fermi-Dirac weight. $p^\alpha$ is the momentum operator along each Cartesian direction $\alpha$. Let $$g_{ij}({\bf k},\omega)=\frac{2\pi e^2}{3m^2\omega \Omega} [f(\epsilon_{i,\bf k})-f(\epsilon_{j,\bf k})]\delta(\epsilon_{j,{\bf k}}-\epsilon_{i,{\bf k}}-\hbar \omega).$$ Then suppressing the explicit dependence of $\sigma$ on \textbf{k} and $\omega$, the conductivity can be expressed as:
\begin{equation}
\begin{aligned}
\sigma=\sum_{i,j,\alpha} g_{ij}\int d^3x\int d^3x^\prime[\psi_{j}^*(x)p^\alpha\psi_i(x)][\psi_{i}^*(x^\prime)p^\alpha\psi_j(x^\prime)],
\label{mom_oper_expansion}
\end{aligned}
\end{equation}
a form that reminds of the the current-current correlation function origins of Kubo's approach. If we define complex valued functions $\xi_{ij}^\alpha(x)=\psi_i^*(x)p^\alpha\psi_j(x)$ on a real space grid (call them \textbf{x}) with uniform spacing of width \textit{h} in three dimensions, then we can approximate the integrals as a sum on the grid. Thus, Eq. (\ref{mom_oper_expansion}) can be written as: 
\begin{equation}
\sigma \approx h^6\sum_{x,x^\prime}\sum_{i,j,\alpha} g_{ij}\xi_{ji}^\alpha(x)\xi_{ij}^\alpha(x^\prime)
\label{sigma_approx}
\end{equation}
In the preceding, the approximation becomes exact as $h \rightarrow 0$. If we define a Hermitian, positive-semidefinite matrix: 
\begin{equation}
\Gamma(x,x^\prime) = \sum_{i,j,\alpha} g_{ij}\xi_{ji}^\alpha(x)\xi_{ij}^\alpha(x^\prime)
\label{gamma}
\end{equation}we can spatially decompose the conductivity at each grid point as $\zeta(x) = \mid\sum_{x^\prime}\Gamma(x,x^\prime)\mid$. $\zeta(x)$ contains vital information about the conduction-active parts of the system\footnote{A simpler scheme is to just look at the structure of the Kohn-Sham eigenfunctions near the Fermi level to identify the conduction active parts of the network. While this is a sensible first approximation, it entirely neglects the current-current correlations that underlie the derivation of The Kubo formula from linear response theory.}.

To implement the method, we used VASP and associated Kohn-Sham orbitals $\psi_{i,\bf k}$. We divided the supercell into $36\times 36\times 36\ (dim \Gamma = 46656)$ grid points and obtained the wavefunction at each point by using the convenient code of R. M. Feenstra and M. Widom \cite{widom}. In computing the $\xi_{ij}^\alpha$, we used a centered finite-difference method to compute the gradient of $\psi_i$ for each $\alpha$. We used an electronic temperature of T = 1000 K for the Fermi-Dirac distribution. We approximated the $\delta$ function in Eq. (\ref{kgf_eqn}) by Gaussian distribution of width \textit{k}T, where \textit{k} is Boltzmann's constant.

\section{\label{sec:level3}Results}
\subsection{\label{sec:level3_a}Bonding and topology of the models}
\begin{figure}[!h]
\begin{center}
\includegraphics[width=1.65in]{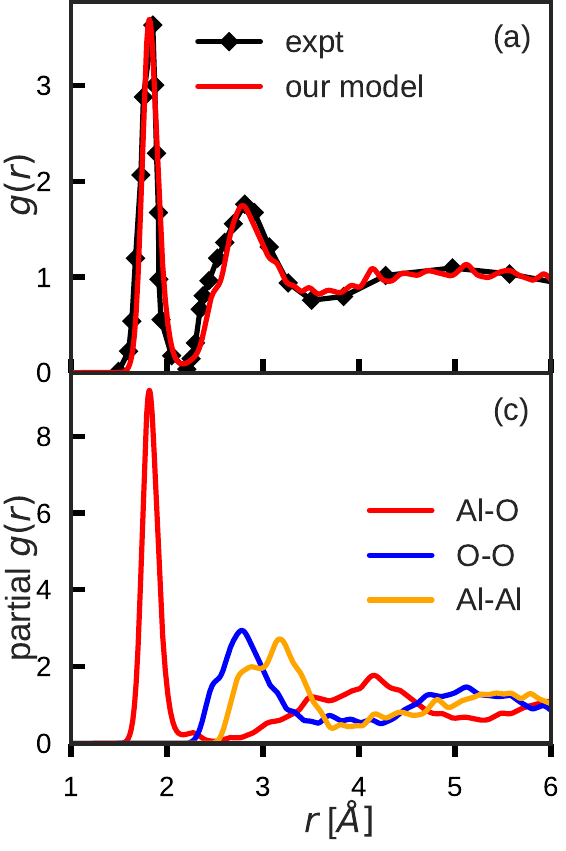}
\includegraphics[width=1.65in]{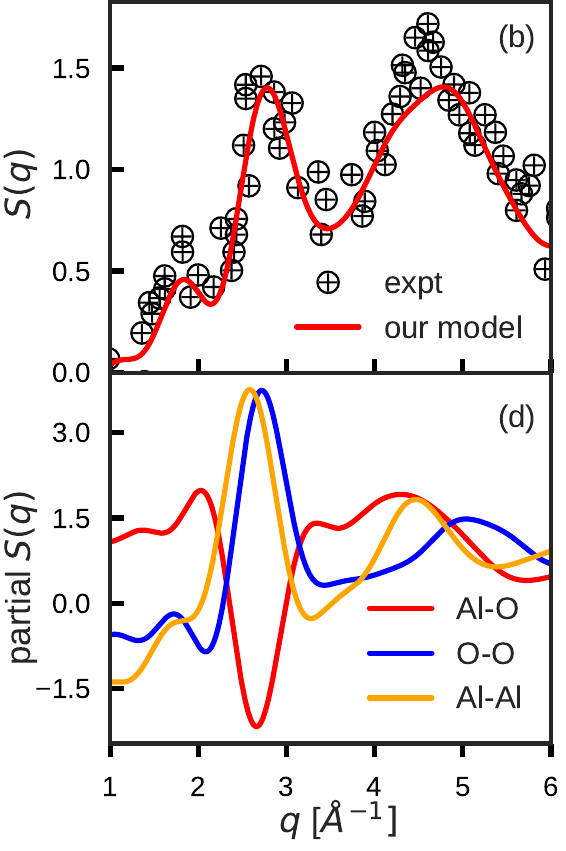}
\caption{$g(r)$ and $S(q)$ of a-Al$_2$O$_3$:~(a) $g(r)$ computed on models are compared with measured $g(r)$. (b) $S(q)$ computed on models are compared with measured $S(q)$. (c) and (d) partial $g(r)$ and partial $S(q)$ respectively for Al-O, O-O and Al-Al pairs.}
\label{rdf_exp}
\end{center}
\end{figure}

\begin{figure}[!h]
     \centering
     \includegraphics[width=3.3in]{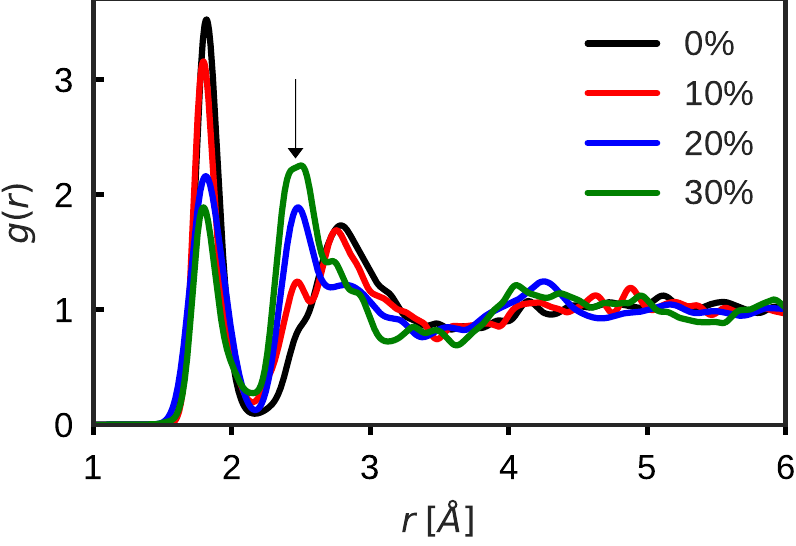}
     \caption{Total $g(r)$ computed from the models of a-Al$_2$O$_3$:Cu at various concentrations of Cu. The hump appearing in Cu-doped models, indicated by arrow, originates from Cu-Cu correlation.}
     \label{rdf_cumodels}
 \end{figure}

\begin{figure}[!h]
 \begin{center}
 \includegraphics[width=3.4in]{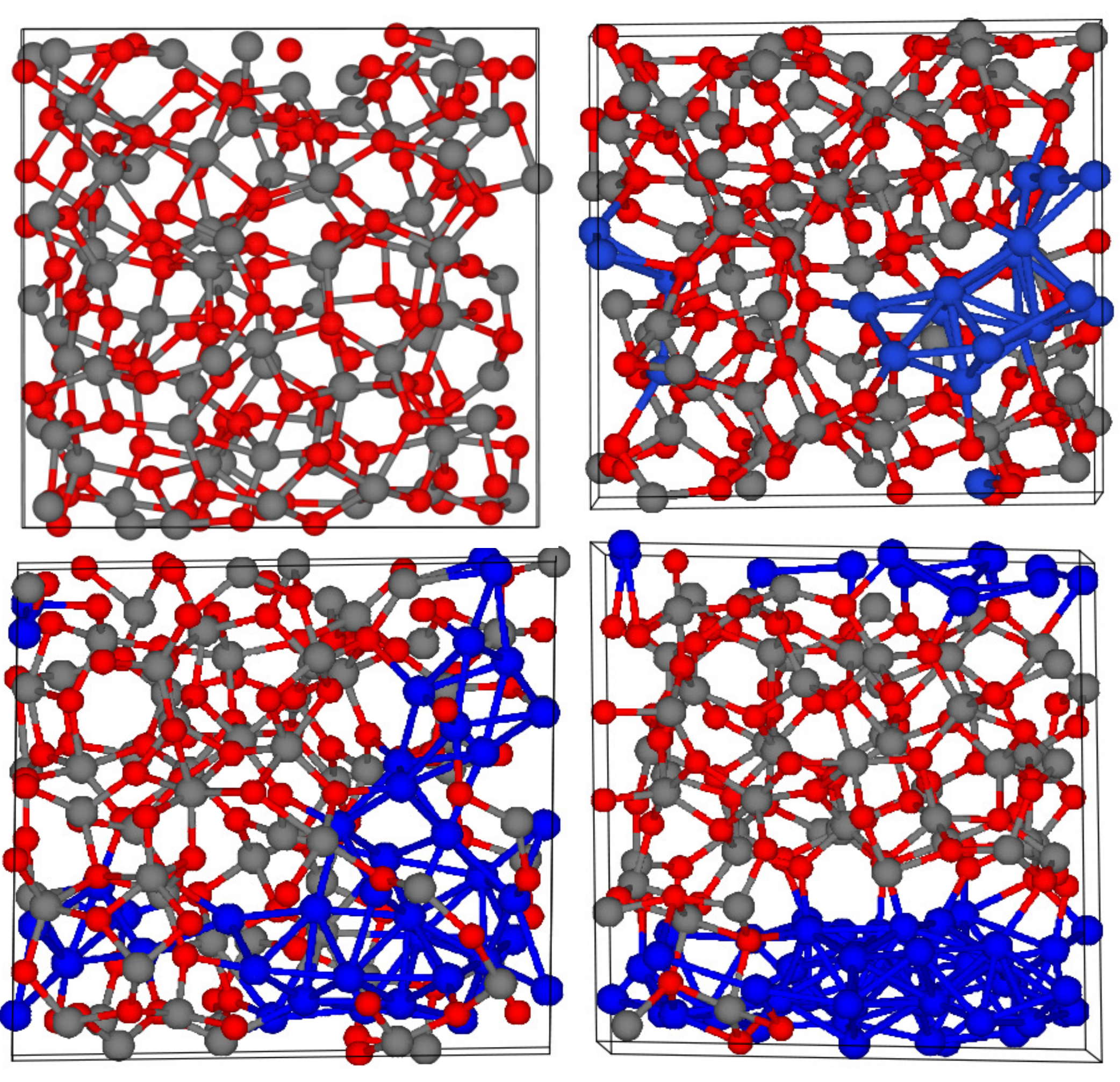}
 \caption{Final relaxed a-Al$_2$O$_3$:Cu models. Top plots (from left) represent for 0\%, 10\% Cu and bottom plots (from left) represent for 20\% and 30\% Cu. Atoms color: Al (gray), Cu (blue) and O (red).}
 \label{structure_cumodels}
 \end{center}
 \end{figure}

As a test of validity of our models, we compute the total radial distribution function, $g(r)$, on a-Al$_2$O$_3$ models and compare with experimentally measured neutron scattering $g(r)$ from \cite{LAMPARTER1997405}. A plot showing these two $g(r)$ is presented in Fig. \ref{rdf_exp} and shows that the models capture the structural order upto 6 \AA~reasonably well. We also compute the structure factor, $S(q)$, on our models at 2600 K and compare it with $S(q)$ measured on l-Al$_2$O$_3$ \cite{landron_liquid_alumina}. The plot shows that these two $S(q)$ show a satisfactory agreement, especially on the positions of peaks at 1.8 \AA$^{-1}$, 2.8 \AA$^{-1}$, 4.7 \AA$^{-1}$. The bottom left plot in Fig. \ref{rdf_exp} presents the partial $g(r)$ computed on models of a-Al$_2$O$_3$. The peaks at 1.81 \AA, 2.78 \AA\ and 3.17 \AA\ correspond to the geometrical bond distances for Al-O, O-O and Al-Al pairs respectively; these results are in agreement with similar earlier works \cite{kummel,momida,sankaran}. The bottom right plot in Fig. \ref{rdf_exp} shows the partial $S(q)$ corresponding to Al-Al, Al-O and O-O pairs computed on a-Al$_2$O$_3$ models. We see that the first peak in the total $S(q)$ occurs at 2.8 \AA$^{-1}$ due to the partial cancellation arising from Al-O correlations.
For doped models, the computed $g(r)$ are plotted in Fig. \ref{rdf_cumodels} and shows that the position of first peak remains largely the same as undoped a-Al$_2$O$_3$ suggesting that Al-O bond remains unaltered. As the concentration of Cu increases, a hump corresponding to Cu-Cu correlation appears and grows at $r\approx$ 2.44 \AA. The relative sharpness of Cu-Cu hump, even for the lowest concentration of Cu, provides a hint that Cu atoms are probably clustered. Indeed, a visual inspection of the models, shown here in Fig. \ref{structure_cumodels}, clearly shows the strong tendency of Cu-atoms to cluster.\\

It is significant that Cu strongly tends to cluster. A study by Dawson and Robertson \cite{dawson_cu_in_al2o3} asserts that the Cu-Cu interactions become more favorable with increasing Cu content. We study the average coordination number around Cu atom at different Cu-concentrations as shown in table \ref{table2}. 
We take the first minima in partial $g(r)$ as the cutoff distance to define the coordination number. The increase in Cu-coordination by Cu and the decrease in Cu-coordination  by Al and O supports the segregation of Cu from the host and formation of cluster. 

 \begin{table}[!htb]
 \caption{Average coordination numbers around Cu atoms for 10\%, 20\% and 30\% Cu models.}
\begin{center}
\begin{tabular*}{0.9\linewidth}{@{\extracolsep\fill}cccc@{\extracolsep\fill}}
\hline
\textbf{Cu content(\%)}&\textbf{Cu-O} &\textbf{Cu-Cu}&\textbf{Cu-Al}\\
\hline
10 & 1.15 & 5.1 & 3.0  \\
20 & 0.68 & 6.85 & 2.45  \\
30 & 0.48 & 8.27 & 1.78  \\
\hline
\end{tabular*}
\end{center}
\label{table2}
\end{table}

 \subsection{\label{sec:level3_b}Electronic structure}
\subsubsection{\label{sec:level3_b_1}Density of States and the Localization}
\begin{figure}[!h]
\begin{center}
\includegraphics[width=3.4in]{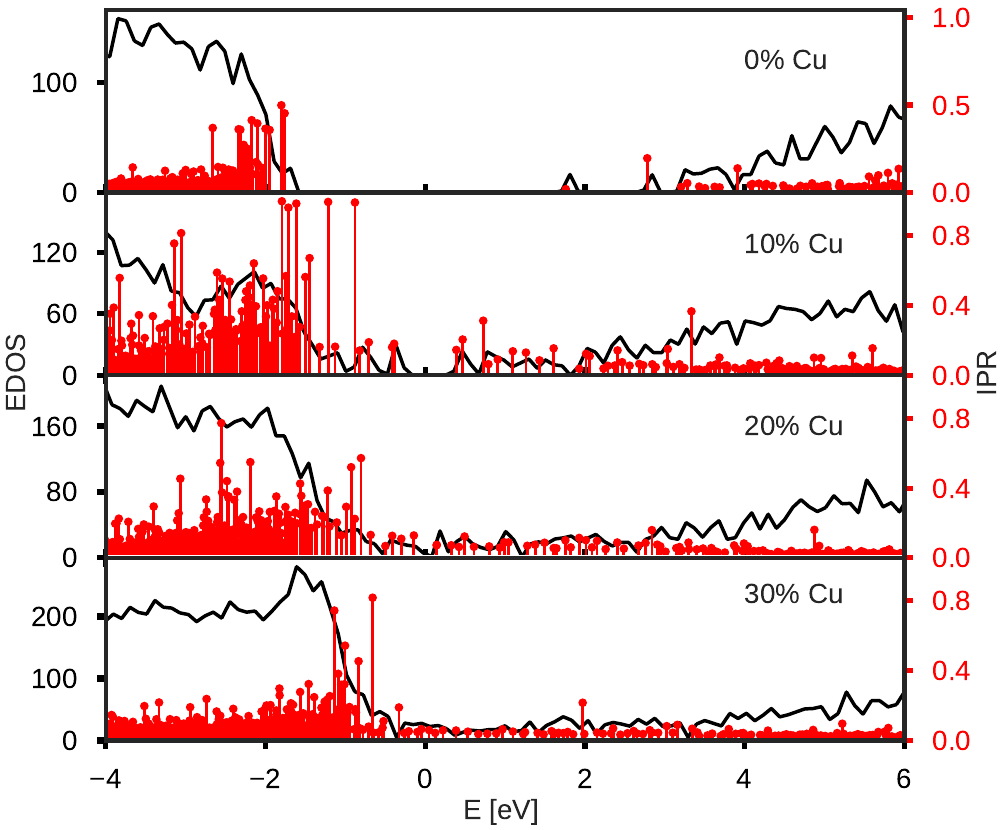}
\caption{Electronic density of states (EDOS) and the inverse participation
 ratio (IPR) computed from a-Al$_2$O$_3$:Cu models for different concentrations of Cu. The black curve represents EDOS and red vertical lines show IPR. The Fermi level is shifted to zero in all plots.}
\label{figiii}
\end{center}
\end{figure}

\begin{figure}[htp]
\begin{center}
\includegraphics[width=3.1in]{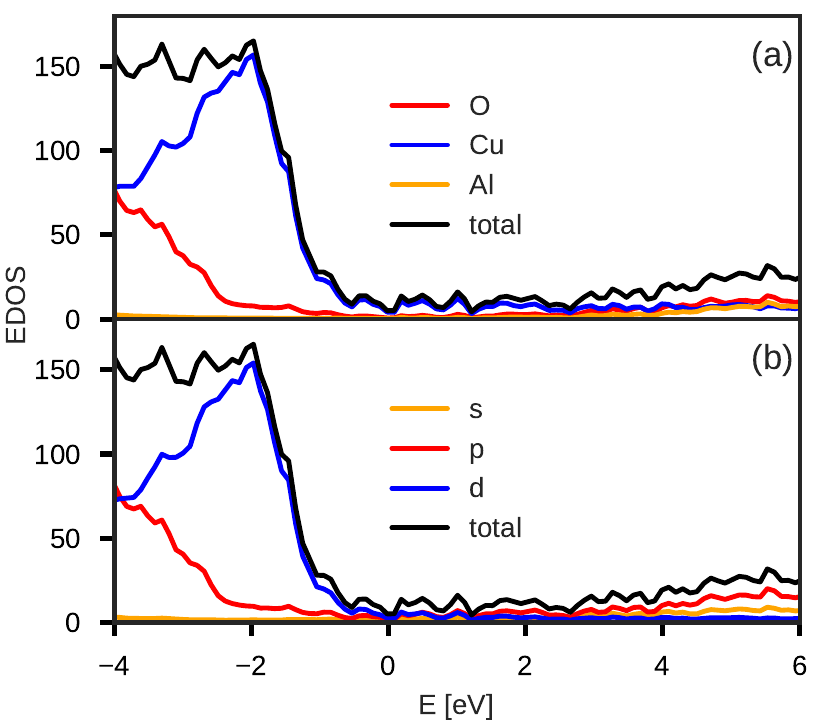}
\caption{Projected electronic density of states (EDOS) computed from a-Al$_2$O$_3$:Cu models with 20\% Cu. (a) Site projected EDOS (b) Orbital projected EDOS. The Fermi energy is shifted to zero.}
\label{figiv}
\end{center}
\end{figure}

Doping by copper in a-Al$_2$O$_3$ is expected to have effects on electronic properties which are of interest for applications of these materials in CBRAM devices. We investigate these effects by examining the density of Kohn-Sham eigenstates (EDOS) and their spatial localization. The localization is gauged by computing the inverse participation ratio (IPR) that is defined as IPR=${\sum_{i}{a_{ni}}^{4}}/{(\sum_{i}{a_{ni}}^{2})^2}$ \cite{ipr}, where the $a_{ni}$'s are the contribution to eigenfunction $\psi_n$ from the $i^{th}$ atomic projected orbital obtained from VASP. Fig. \ref {figiii} shows the computed EDOS and IPR as a function of Cu-concentration. We find a decrease in HOMO-LUMO gap with increasing Cu-concentration; at Cu-concentration 20\% and 30\%, The EDOS is continuous across the Fermi level. The states that fill-in the band gap are quite extended as indicated by small values of IPR around the Fermi level in Fig. \ref {figiii}.
The mean IPR values around the gap declines monotonically with Cu-concentration.\\
\\
By projecting the electronic states onto atomic sites, we observe that the states near the Fermi level for the doped models consist of Cu-orbitals. An example of the site projected EDOS, for 20\% Cu, is plotted in Fig. \ref{figiv}. It is quite interesting that at 20\% and 30\% Cu-concentrations, Cu levels almost {\it uniformly }fill the host a-Al$_2$O$_3$ gap. The Cu does not form an impurity band, as one might naively suppose from experience on heavily-doped semiconductors. We see that models with higher Cu-concentration produce states near Fermi level that yield an essentially metallic form of conduction. This is qualitatively different than the case of Ag in GeSe$_3$ \cite{prasai_gese3}, wherein the Ag atoms do not cluster and do not introduce states in the optical gap of the host. We observe that electron states in the gap are filled mostly by 3d, 4s and 4p orbitals of Cu.
\subsubsection{\label{sec:level3_b_4}Charge analysis on Cu atoms} 
\begin{figure}[t]
 \begin{center}
 \includegraphics[width=3.0in]{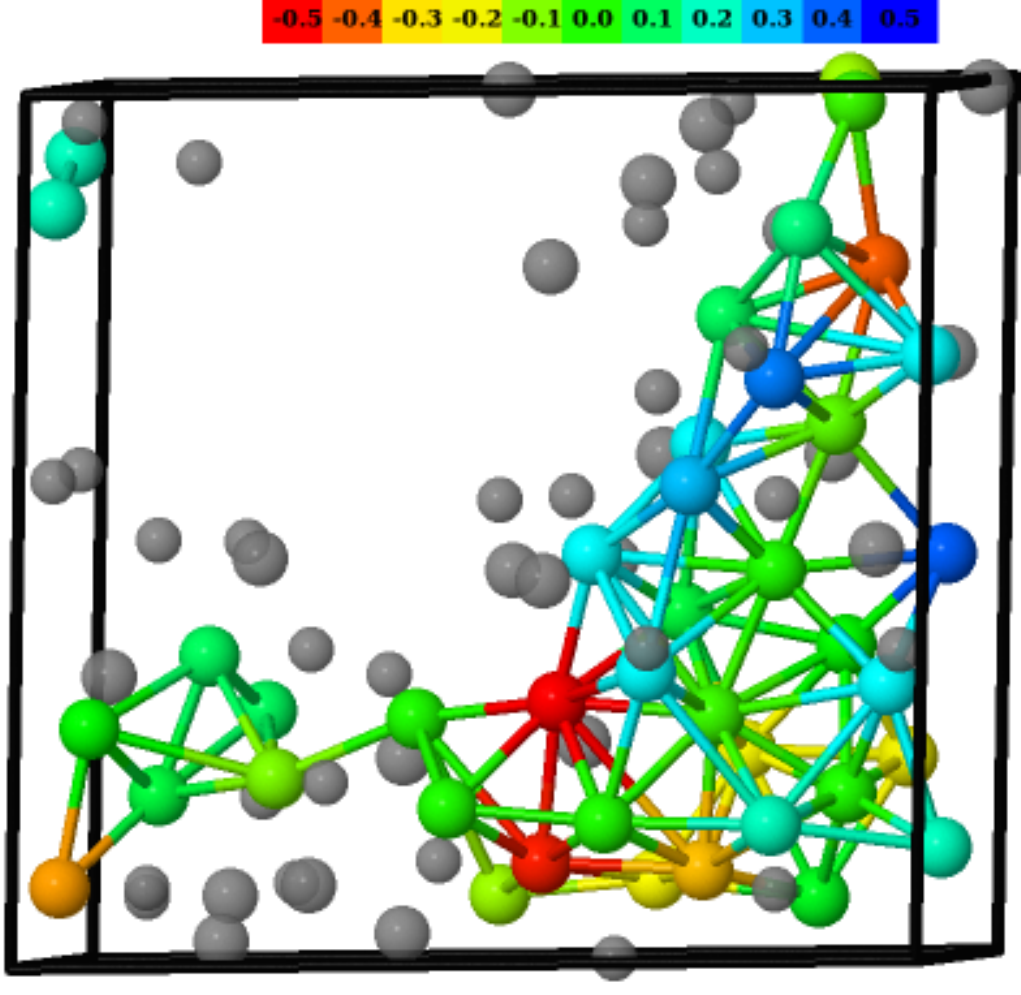}
 \caption{Net Bader charge on Cu atoms calculated from a-Al$_2$O$_3$:Cu models with 20\% Cu-concentration. A color code displayed on top is used to represent the charge state. Charge state of zero, shown by green, represents a neutral Cu atom; the charge values are in units of electronic charge. All Cu-atoms are shown in color. Light gray atoms represent Al and O atoms within the first cutoff distance of Cu atoms.}
 \label{fig_charge_Cu}
 \end{center}
 \end{figure}

The formation of Cu-cluster in a-Al$_2$O$_3$ matrix leaves the Cu atoms in different charge states depending on the local environment of these Cu atoms with O and/or Al atoms. We performed Bader charge analysis \cite{bader_code_ref} to calculate net charge on these atoms and an analysis for 20\% Cu-doped model is shown in  Fig. \ref{fig_charge_Cu}. The charge state of the Cu atoms (shown in color in Fig. \ref{fig_charge_Cu}) can be explained by a simple analysis of the first neighbors around the Cu atoms. Among all the Cu-atoms shown in the figure, only five Cu atoms have exclusively Cu neighbors and are neutral in nature; the rest of the Cu are neighbors with at least one Al or O atoms. When a Cu atom is a neighbor with Al or O atoms, bonding or charge transfer occurs. A Cu atom bonded with O atoms is positively charged, whereas a Cu atom bonded with Al atoms is slightly negatively charged and can be understood in terms of difference in electronegativities of Cu and Al. When a Cu atom is bonded with both O and Al atoms, it is charge neutral. The charge compensation likely to happen in such bonding. The Cu atoms shown in green are therefore almost metallic in nature and are likely to form a conducting channel for the current to flow in the network.

\subsubsection{\label{sec:level3_b_2} Thermally driven conduction fluctuations}
\begin{figure}[!htb]
 \begin{center}
 \includegraphics[width=3.2in]{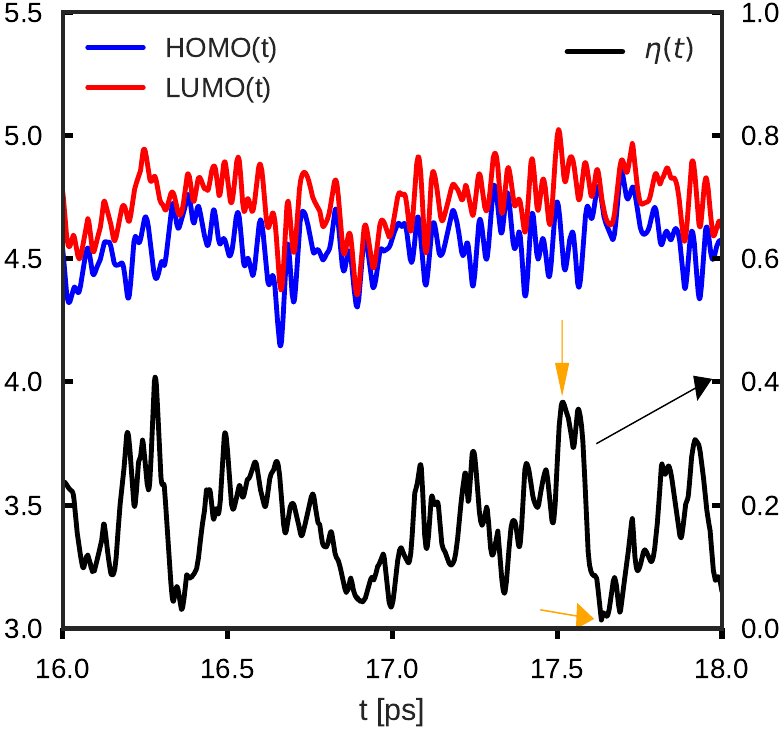} 
 \caption{Fluctuation of HOMO, LUMO and HOMO-LUMO gap ($\eta$) with time for 20\% model at 1000 K. $\eta(t)$ is represented by black line with its values given by right axis of the plot as shown by arrowhead in the plot.}
 \label{figv}
 \end{center}
 \end{figure}

In this section, we discuss relatively dramatic thermally-induced fluctuations in the HOMO-LUMO splitting and consider the electronic conduction mechanisms\footnote{Here and elsewhere in this paper, electronic time evolution refers only to variation in Kohn-Sham eigenvalues/states on the Born-Oppenheimer surface -- no attempt is made to solve a time-dependent Kohn-Sham equation}. We illustrate with one of the conducting models (including 20\% Cu) and performed molecular dynamics (MD) at 1000 K for 24 ps. The fluctuation of the frontier HOMO and LUMO levels with time is provided in Fig. \ref{figv}. $\eta(t)$ is the HOMO-LUMO splitting through the course of the MD. The model reveals a large thermally driven fluctuation in the value of the HOMO-LUMO gap with time.

 To physically interpret the connection of the gap with electronic conductivity ($\sigma$), let us write a simple expression for the dc conductivity (T = 0 K) following Mott and Davis \cite{mott_davis},
\begin{equation}
\sigma_{dc} = \frac{2\pi e^2 \hbar \Omega}{m^2}{\mid D_{\epsilon_f} \mid}^2 N^2({\epsilon_f})
\label{kgf_fermi}
\end{equation}

where $D_{\epsilon_f}$ is a matrix element of $\nabla_\alpha$ between Kohn-Sham states near the Fermi level and $N$($\epsilon_f$) is the density of states. So, for dc conduction to occur, there needs to be finite density of states at the Fermi level (to enable electronic transitions, as from Fermi's Golden Rule) and non-vanishing matrix elements ${\mid D_{\epsilon_f}\mid}^2$ as in Eq. (\ref{kgf_eqn}). We expect more available states near the Fermi level for the system with small gap, thus the conductivity $\sigma (t)$ can be very crudely linked to $\eta(t)$ (small $\eta \implies$ large $\sigma$) in the spirit of Landau-Zener tunneling \cite{landau,zener}. We provisionally interpret the small gap (small $\eta$) instantaneous configurations as low resistance states, and the large gap configurations as high resistance states.\\

It is therefore interesting to visualize the conduction-active parts of the network for these different states. We selected two snapshots (shown by orange arrows in Fig. \ref{figv}), one representing a small gap (low $\eta$) and the other large gap (high $\eta$) from the simulation and obtained the SPC as described in section \ref{sec:level2_b}. The variation of the HOMO-LUMO gap due to thermal fluctuations has also been studied in Boron-doped $a$-Si at 600 K, where it was observed that with addition of hydrogen to the network, there occurs a thermal modulation of HOMO and LUMO states causing the HOMO and LUMO states to be overlapped at a certain interval of the thermal simulation representing highly conducting configuration \cite{anup_B-Si}. This computation makes it clear that the DC conductivity is difficult to accurately estimate, since to handle the large electron-phonon coupling for states near the Fermi level, long MD averages at constant temperature would be required (within an adiabatic picture for which one simply averages the Kubo formula over a trajectory. \\

 \subsubsection{\label{sec:level3_b_3} Space-Projected Conductivity}
\begin{figure}[!htb]
 \centering 
  \includegraphics[width=3.4in]{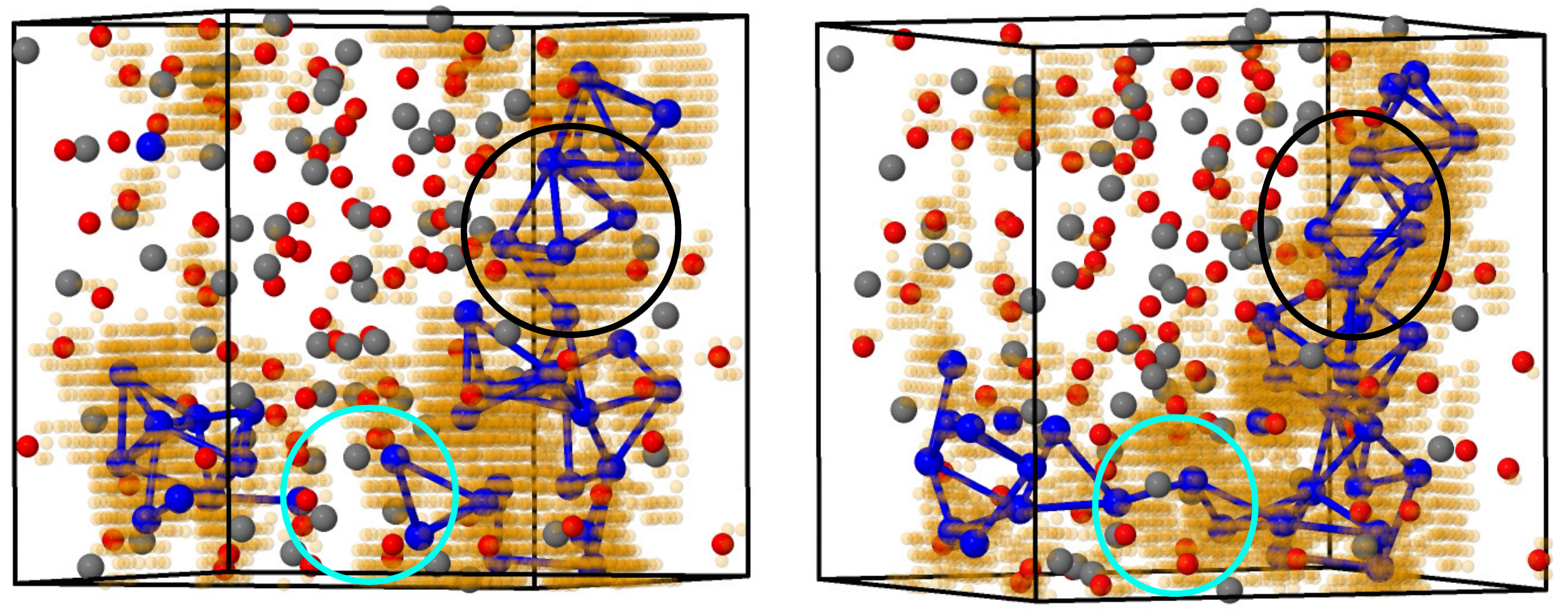}
 \caption{Overlaying SPC values with atomic configuration: On the left, large gap (high resistance) state of a-Al$_2$O$_3$:Cu model with 20\% Cu. On the right, small gap (low resistance) state of the same system. Color nomenclature: blue- Cu atom, red- O atom and gray- Al atom. The bond length of cutoff 2.6 \AA\ is chosen. Circles with same color represent same part of local configurations. There is a factor of about $10^4$ between the conductivities of the two conformations.}
 \label{isosurface}
 \end{figure}

 We investigated SPC by computing $\zeta(x)$ as described in section \ref{sec:level2_b} in our models. SPC values are evaluated at coarse 3D grid  points inside the supercells. A graphical representation of SPC values in 3D grid points overlaid with the atomic configuration is shown in Fig. \ref{isosurface}. This figure shows the SPC computed on two models: one with large $\eta$ and the other with and small $\eta$. We include 12\% of the highest local contributions to SPC in each plot. The SPC reveals that the conduction path is primarily along interconnected Cu atoms. A few O atoms in the vicinity of Cu atoms also participate in the conduction whereas Al atoms do not show any role in the conduction. We see that the SPC for the large gap snapshot is disconnected so that $\zeta(x)$ appears to be localized in certain region whereas the SPC with small gap forms an interconnected chain for the conduction. For these two particular structures, we observed the local configurations as shown by the enclosed circles of Fig. \ref{isosurface} where the Cu atoms come closer to form short bonds and form a closed network. This shows that the connectivity among Cu atoms determines the conductivity of the system. Besides the structural difference, the type and the number of clusters also affect the HOMO-LUMO gap. It has been shown that an alternation of the HOMO-LUMO gap occurs between even and odd numbered isolated clusters due to electron-pairing effects and particularly large gap for cluster size 2, 8, 18, 20, 34 and 40 which are also called as magic clusters \cite{kabir_cu_clusters}. At this temperature, the diffusion of Cu atoms may cause the change in the bonding environment of Cu atoms resulting in the variation of the gap with time.\\
\subsection{\label{sec:level3_c}Ionic motion}
As a representative example, the 20\% model was annealed at different temperatures 800 K and 1000 K for 15 ps, and the resulting ion dynamics were studied by calculating the mean-squared displacement for each atomic species as: 

\begin{equation}
\begin{aligned}
\langle {{r^2}(t)}\rangle _\alpha = \frac{1}{N_a}\sum_{i}^{N_\alpha} \langle {\mid\vec{r_i}(t)-\vec{r_i}(0)\mid}^2 \rangle
\end{aligned}
\end{equation}

 where $N_\alpha$ represents the number of atoms of species $\alpha$, $r_i(t)$ represents the position of atom $i$ at time t, and the $\langle$ $\rangle$ represents an average on the time steps and/or the particles. The connection between mean-squared displacement and the self-diffusion coefficient is given by Einstein's relation
\begin{equation}
\centering
\langle {{r^2}(t)}\rangle = A + 6Dt
\label{einstein_eqn}
\end{equation}
where $D$ is the self-diffusion coefficient, $A $ is a constant and $t$ is the simulation time.
\begin{figure}[!h]
 \begin{center}
 \includegraphics[width=3.2in]{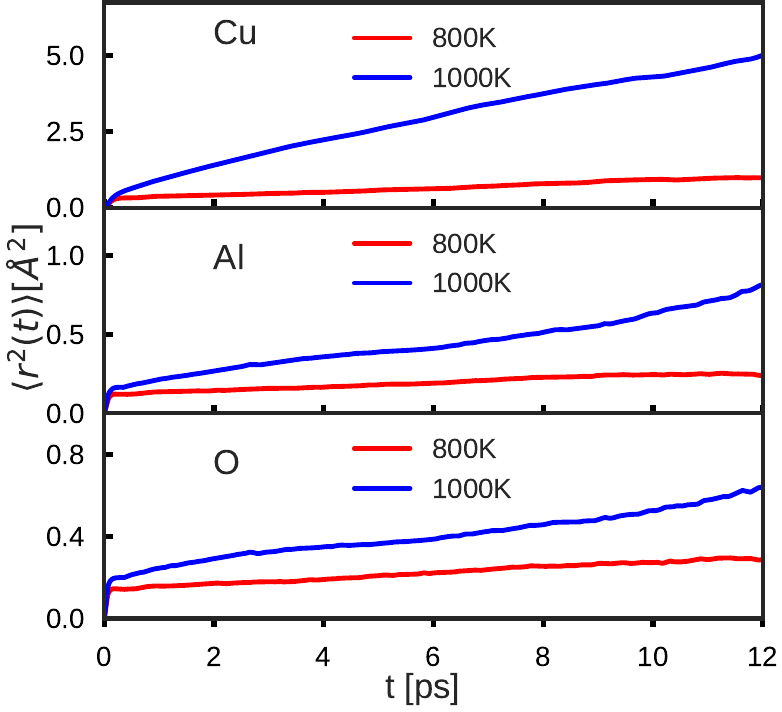}
 \caption{Mean-squared displacement at 800 K and 1000 K for 20\% Cu model.}
 \label{figviii}
 \end{center}
 \end{figure}
Figure~\ref{figviii} shows the mean-squared displacement for the corresponding species. Clearly, Cu atoms are more diffusive than Al and O atoms. On taking the snapshots of the position of atoms (figures not shown here), we find that the Cu atoms do not diffuse into the host matrix but diffuse within the Cu clusters and thus the Cu clusters become stable at these range of temperatures. We then calculated the self-diffusion coefficient for each species using Eq. (\ref{einstein_eqn}). The diffusion coefficient for Cu at 800 K and 1000 K are obtained to be $9.95 \times 10^{-7} cm{^2}/s$ and $6.248 \times 10^{-6} cm{^2}/s$ respectively. Cu is relatively static in a-Al$_2$O$_3$ compared to chalcogenides \cite{binaya_chalcogen}.
\subsection{\label{sec:level3_d}Lattice Dynamics}
 We study the lattice dynamics of these Cu-doped systems by the means of vibrational density of states (VDOS), species projected VDOS and the vibrational IPR. The properties are studied within the harmonic approximation using the first principles method. The dynamical matrix is obtained by displacing each atoms by 0.015~\AA\ along $\pm x$, $\pm y$ and $\pm z$ directions. The diagonalization of the dynamical matrix yields eigenfrequencies and the corresponding eigenmodes. The normalized VDOS and the partial VDOS are expressed as \cite{pasquarello}
\begin{align}
Z(E) & =\frac{1} {3N}\sum_{n}\delta(E-{\hbar \omega_{n}})
\label{eqn:7}\\
Z_\alpha(E) &=\frac{1}{3N}\sum_{i\in\alpha}^{N_\alpha}\sum_{n}{\mid e_{i}^{n}\mid}^2\delta(E-{\hbar \omega_{n}})\label{eq:8}
\end{align}
where $\omega_n$ are the normalized eigenfrequencies (3N in total). Here, the sum over $i$ is over all the atoms belonging to the species $\alpha$ and $e_{i}^{n}$ corresponds to the displacement vector of atom $i$ with Cartesian components $e_{i\mu}^{n}$ where $\mu$ = x, y and z. We approximate the $\delta$ function by a Gaussian distribution function of width 10 $cm^{-1}$. Among the 3N eigenmodes, we neglect the first three translational modes with frequency very close to zero. \\
 \begin{figure}[!h]
 \begin{center}
 \includegraphics[width=3.4in]{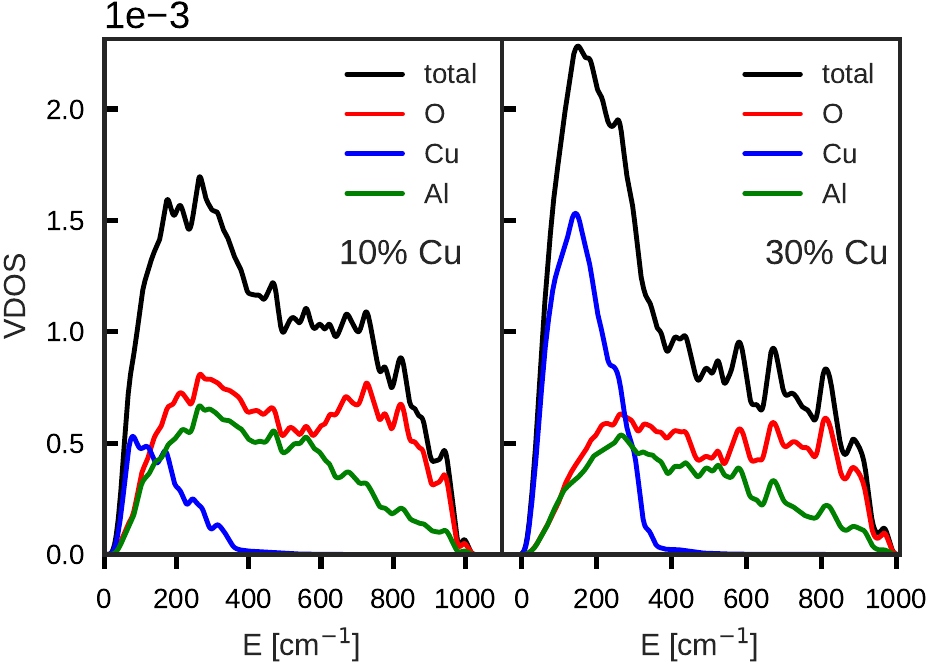}
 \caption{Normalized total and partial vibrational density of states for 10\% and 30\% of Cu models.}
 \label{figix}
 \end{center}
 \end{figure}
 \begin{figure}[!h]
\begin{center}
\includegraphics[width=3.4in]{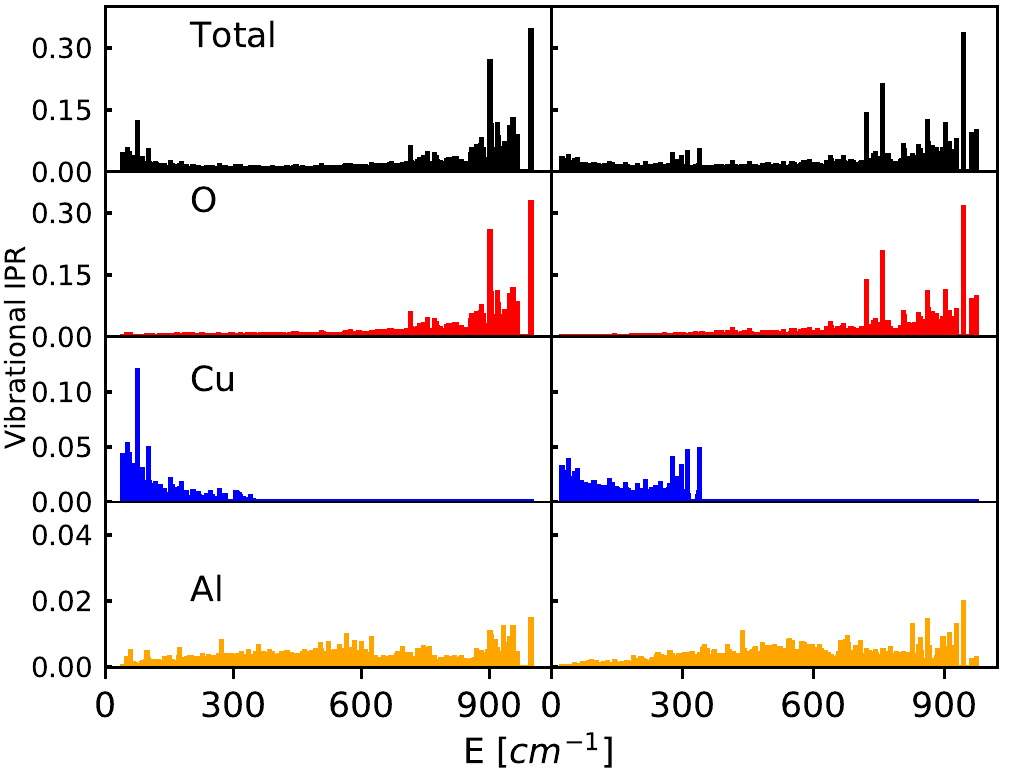} 
 \caption{Vibrational IPR different models. Left column for 10\% model and the right column for 30\% model.}
 \label{figx}
 \end{center}
 \end{figure}
 Figure \ref{figix} shows the total and partial VDOS for 10\% and 30\% Cu content. The lower vibrational modes correspond to the Cu atoms. The higher frequency modes are unsurprisingly dominated by O atoms. To study the localization of the vibrational eigenstates, we calculated the vibrational IPR for each species. From Fig. \ref{figx}, we see that the higher modes corresponding to the O atoms are more localized compared to the lower modes for both concentrations of Cu. The lower eigenstates corresponding to Cu for 10\% Cu model are quite localized compared to the 30\% Cu model. The vibrational states for aluminum are mostly extended for both models. \\
\section{Conclusion}
In this paper, we studied realistic models of a-Al$_2$O$_3$:Cu, and showed that the Cu atoms have a strong propensity to cluster in the ionic a-Al$_2$O$_3$ host. We observed a continuous filling of the optical gap by Cu levels, especially at 20\% and 30\% models. As the Cu-concentration increases (and Cu-Cu connectivity increases), the Cu levels band to enable metallic conduction. We observed the opening and closing of the HOMO-LUMO gap at an elevated temperature, and projected electronic conductivity into real space and visualized the conduction-active parts of the network. We showed that the connectivity of Cu atoms play a significant role in the electronic conduction. We studied the diffusion of Cu atoms in a-Al$_2$O$_3$ at different temperatures and observed that the Cu atoms do not diffuse easily into the a-Al$_2$O$_3$ in contrast with relatively covalent chalcogenides like GeSe$_3$ \cite{binaya_chalcogen}. We discussed the harmonic lattice dynamics of the models by calculating vibrational density of states and the vibrational IPR and showed that the lower vibrational modes correspond to Cu atoms and the higher modes correspond to O atoms.\\
\\
The results presented in this work on a-Al$_2$O$_3$:Cu show an interesting contrast with similar study performed on GeSe$_3$:Ag \cite{ks_2019}. We find that the properties of Cu in the oxide host (in this case, a-Al$_2$O$_3$:Cu) contrast with that of Ag in chalcogenide (in case of \cite{ks_2019}, GeSe$_3$:Ag). The Ag atoms do not form a cluster in the GeSe$_3$ and no uniform filling of the optical gap is observed. In other words, one has to electrochemically work hard to draw Ag atoms together to form a cluster in GeSe$_3$. So, the electronic conduction is likely to occur by hopping process in GeSe$_3$:Ag whereas the conduction in Al$_2$O$_3$:Cu is most likely through the interconnected Cu atoms in the network. We observed that Cu in a-Al$_2$O$_3$ exhibits different charge states (negative, neutral and positive) whereas the charge state of Ag in GeSe$_3$ changes from neutral when isolated to ionic (positive) near the trapping center sites (host atoms) \cite{choudary_Ag_dynamics}.
\section{Acknowledgments}
 We thank US NSF support under grants DMR 1507670 and 1506836. We thank NVIDIA Corporation for donating a Tesla K40 GPU which was used in some of our calculations. Some of this work used the Extreme Science and Engineering Discovery Environment (XSEDE), which is supported by National Science Foundation grant number ACI-1548562, using BRIDGES at the Pittsburgh Supercomputer Center under the allocation TG-DMR180083. We thank Prof. Gang Chen for valuable discussions during this work.\\
\\

\begin{thebibliography}{49}%
\makeatletter
\providecommand \@ifxundefined [1]{%
 \@ifx{#1\undefined}
}%
\providecommand \@ifnum [1]{%
 \ifnum #1\expandafter \@firstoftwo
 \else \expandafter \@secondoftwo
 \fi
}%
\providecommand \@ifx [1]{%
 \ifx #1\expandafter \@firstoftwo
 \else \expandafter \@secondoftwo
 \fi
}%
\providecommand \natexlab [1]{#1}%
\providecommand \enquote  [1]{``#1''}%
\providecommand \bibnamefont  [1]{#1}%
\providecommand \bibfnamefont [1]{#1}%
\providecommand \citenamefont [1]{#1}%
\providecommand \href@noop [0]{\@secondoftwo}%
\providecommand \href [0]{\begingroup \@sanitize@url \@href}%
\providecommand \@href[1]{\@@startlink{#1}\@@href}%
\providecommand \@@href[1]{\endgroup#1\@@endlink}%
\providecommand \@sanitize@url [0]{\catcode `\\12\catcode `\$12\catcode
  `\&12\catcode `\#12\catcode `\^12\catcode `\_12\catcode `\%12\relax}%
\providecommand \@@startlink[1]{}%
\providecommand \@@endlink[0]{}%
\providecommand \url  [0]{\begingroup\@sanitize@url \@url }%
\providecommand \@url [1]{\endgroup\@href {#1}{\urlprefix }}%
\providecommand \urlprefix  [0]{URL }%
\providecommand \Eprint [0]{\href }%
\providecommand \doibase [0]{http://dx.doi.org/}%
\providecommand \selectlanguage [0]{\@gobble}%
\providecommand \bibinfo  [0]{\@secondoftwo}%
\providecommand \bibfield  [0]{\@secondoftwo}%
\providecommand \translation [1]{[#1]}%
\providecommand \BibitemOpen [0]{}%
\providecommand \bibitemStop [0]{}%
\providecommand \bibitemNoStop [0]{.\EOS\space}%
\providecommand \EOS [0]{\spacefactor3000\relax}%
\providecommand \BibitemShut  [1]{\csname bibitem#1\endcsname}%
\let\auto@bib@innerbib\@empty
\bibitem [{\citenamefont {Simmons}\ and\ \citenamefont
  {Verderber}(1967)}]{history1}%
  \BibitemOpen
  \bibfield  {author} {\bibinfo {author} {\bibfnamefont {J.~G.}\ \bibnamefont
  {Simmons}}\ and\ \bibinfo {author} {\bibfnamefont {R.~R.}\ \bibnamefont
  {Verderber}},\ }\href@noop {} {\bibfield  {journal} {\bibinfo  {journal}
  {Radio and Electronic Engineer}\ }\textbf {\bibinfo {volume} {34}},\ \bibinfo
  {pages} {81} (\bibinfo {year} {1967})}\BibitemShut {NoStop}%
\bibitem [{\citenamefont {Waser}\ \emph {et~al.}(2009)\citenamefont {Waser},
  \citenamefont {Dittmann}, \citenamefont {Staikov},\ and\ \citenamefont
  {Szot}}]{history2}%
  \BibitemOpen
  \bibfield  {author} {\bibinfo {author} {\bibfnamefont {R.}~\bibnamefont
  {Waser}}, \bibinfo {author} {\bibfnamefont {R.}~\bibnamefont {Dittmann}},
  \bibinfo {author} {\bibfnamefont {G.}~\bibnamefont {Staikov}}, \ and\
  \bibinfo {author} {\bibfnamefont {K.}~\bibnamefont {Szot}},\ }\href@noop {}
  {\bibfield  {journal} {\bibinfo  {journal} {Advanced Materials}\ }\textbf
  {\bibinfo {volume} {21}},\ \bibinfo {pages} {2632} (\bibinfo {year}
  {2009})}\BibitemShut {NoStop}%
\bibitem [{\citenamefont {Tappertzhofen}\ \emph {et~al.}(2013)\citenamefont
  {Tappertzhofen}, \citenamefont {Valov}, \citenamefont {Tsuruoka},
  \citenamefont {Hasegawa}, \citenamefont {Waser},\ and\ \citenamefont
  {Aono}}]{Valov_2013}%
  \BibitemOpen
  \bibfield  {author} {\bibinfo {author} {\bibfnamefont {S.}~\bibnamefont
  {Tappertzhofen}}, \bibinfo {author} {\bibfnamefont {I.}~\bibnamefont
  {Valov}}, \bibinfo {author} {\bibfnamefont {T.}~\bibnamefont {Tsuruoka}},
  \bibinfo {author} {\bibfnamefont {T.}~\bibnamefont {Hasegawa}}, \bibinfo
  {author} {\bibfnamefont {R.}~\bibnamefont {Waser}}, \ and\ \bibinfo {author}
  {\bibfnamefont {M.}~\bibnamefont {Aono}},\ }\href@noop {} {\bibfield
  {journal} {\bibinfo  {journal} {ACS Nano}\ }\textbf {\bibinfo {volume} {7}},\
  \bibinfo {pages} {6396} (\bibinfo {year} {2013})}\BibitemShut {NoStop}%
\bibitem [{\citenamefont {Kozicki}\ and\ \citenamefont
  {Barnaby}(2016)}]{kozicki_review}%
  \BibitemOpen
  \bibfield  {author} {\bibinfo {author} {\bibfnamefont {M.~N.}\ \bibnamefont
  {Kozicki}}\ and\ \bibinfo {author} {\bibfnamefont {H.~J.}\ \bibnamefont
  {Barnaby}},\ }\href@noop {} {\bibfield  {journal} {\bibinfo  {journal}
  {Semiconductor Science and Technology}\ }\textbf {\bibinfo {volume} {31}},\
  \bibinfo {pages} {113001} (\bibinfo {year} {2016})}\BibitemShut {NoStop}%
\bibitem [{\citenamefont {Dietrich}\ \emph {et~al.}(2007)\citenamefont
  {Dietrich}, \citenamefont {Angerbauer}, \citenamefont {Ivanov}, \citenamefont
  {Gogl}, \citenamefont {Hoenigschmid}, \citenamefont {Kund}, \citenamefont
  {Liaw}, \citenamefont {Markert}, \citenamefont {Symanczyk}, \citenamefont
  {Altimime}, \citenamefont {Bournat},\ and\ \citenamefont
  {Mueller}}]{cbram_stefan}%
  \BibitemOpen
  \bibfield  {author} {\bibinfo {author} {\bibfnamefont {S.}~\bibnamefont
  {Dietrich}}, \bibinfo {author} {\bibfnamefont {M.}~\bibnamefont
  {Angerbauer}}, \bibinfo {author} {\bibfnamefont {M.}~\bibnamefont {Ivanov}},
  \bibinfo {author} {\bibfnamefont {D.}~\bibnamefont {Gogl}}, \bibinfo {author}
  {\bibfnamefont {H.}~\bibnamefont {Hoenigschmid}}, \bibinfo {author}
  {\bibfnamefont {M.}~\bibnamefont {Kund}}, \bibinfo {author} {\bibfnamefont
  {C.}~\bibnamefont {Liaw}}, \bibinfo {author} {\bibfnamefont {M.}~\bibnamefont
  {Markert}}, \bibinfo {author} {\bibfnamefont {R.}~\bibnamefont {Symanczyk}},
  \bibinfo {author} {\bibfnamefont {L.}~\bibnamefont {Altimime}}, \bibinfo
  {author} {\bibfnamefont {S.}~\bibnamefont {Bournat}}, \ and\ \bibinfo
  {author} {\bibfnamefont {G.}~\bibnamefont {Mueller}},\ }\href@noop {}
  {\bibfield  {journal} {\bibinfo  {journal} {IEEE Journal of Solid-State
  Circuits}\ }\textbf {\bibinfo {volume} {42}},\ \bibinfo {pages} {839}
  (\bibinfo {year} {2007})}\BibitemShut {NoStop}%
\bibitem [{\citenamefont {Kund}\ \emph {et~al.}(2005)\citenamefont {Kund},
  \citenamefont {Beitel}, \citenamefont {Pinnow}, \citenamefont {Rohr},
  \citenamefont {Schumann}, \citenamefont {Symanczyk}, \citenamefont {Ufert},\
  and\ \citenamefont {Muller}}]{chalco1}%
  \BibitemOpen
  \bibfield  {author} {\bibinfo {author} {\bibfnamefont {M.}~\bibnamefont
  {Kund}}, \bibinfo {author} {\bibfnamefont {G.}~\bibnamefont {Beitel}},
  \bibinfo {author} {\bibfnamefont {C.~.}\ \bibnamefont {Pinnow}}, \bibinfo
  {author} {\bibfnamefont {T.}~\bibnamefont {Rohr}}, \bibinfo {author}
  {\bibfnamefont {J.}~\bibnamefont {Schumann}}, \bibinfo {author}
  {\bibfnamefont {R.}~\bibnamefont {Symanczyk}}, \bibinfo {author}
  {\bibfnamefont {K.}~\bibnamefont {Ufert}}, \ and\ \bibinfo {author}
  {\bibfnamefont {G.}~\bibnamefont {Muller}},\ }in\ \href@noop {} {\emph
  {\bibinfo {booktitle} {IEEE InternationalElectron Devices Meeting, 2005. IEDM
  Technical Digest.}}}\ (\bibinfo {year} {2005})\ pp.\ \bibinfo {pages}
  {754--757}\BibitemShut {NoStop}%
\bibitem [{\citenamefont {Lee}\ \emph {et~al.}(2017)\citenamefont {Lee},
  \citenamefont {Oukassi}, \citenamefont {Molas}, \citenamefont {Carabasse},
  \citenamefont {Salot},\ and\ \citenamefont {Perniola}}]{chalco2_gese2}%
  \BibitemOpen
  \bibfield  {author} {\bibinfo {author} {\bibfnamefont {D.}~\bibnamefont
  {Lee}}, \bibinfo {author} {\bibfnamefont {S.}~\bibnamefont {Oukassi}},
  \bibinfo {author} {\bibfnamefont {G.}~\bibnamefont {Molas}}, \bibinfo
  {author} {\bibfnamefont {C.}~\bibnamefont {Carabasse}}, \bibinfo {author}
  {\bibfnamefont {R.}~\bibnamefont {Salot}}, \ and\ \bibinfo {author}
  {\bibfnamefont {L.}~\bibnamefont {Perniola}},\ }\href@noop {} {\bibfield
  {journal} {\bibinfo  {journal} {IEEE Journal of the Electron Devices
  Society}\ }\textbf {\bibinfo {volume} {5}},\ \bibinfo {pages} {283} (\bibinfo
  {year} {2017})}\BibitemShut {NoStop}%
\bibitem [{\citenamefont {Chen}\ \emph {et~al.}(2017)\citenamefont {Chen},
  \citenamefont {Tappertzhofen}, \citenamefont {Barnaby},\ and\ \citenamefont
  {Kozicki}}]{chen_sio2}%
  \BibitemOpen
  \bibfield  {author} {\bibinfo {author} {\bibfnamefont {W.}~\bibnamefont
  {Chen}}, \bibinfo {author} {\bibfnamefont {S.}~\bibnamefont {Tappertzhofen}},
  \bibinfo {author} {\bibfnamefont {H.~J.}\ \bibnamefont {Barnaby}}, \ and\
  \bibinfo {author} {\bibfnamefont {M.~N.}\ \bibnamefont {Kozicki}},\
  }\href@noop {} {\bibfield  {journal} {\bibinfo  {journal} {Journal of
  Electroceramics}\ }\textbf {\bibinfo {volume} {39}},\ \bibinfo {pages} {109}
  (\bibinfo {year} {2017})}\BibitemShut {NoStop}%
\bibitem [{\citenamefont {Tsuruoka}\ \emph {et~al.}(2010)\citenamefont
  {Tsuruoka}, \citenamefont {Terabe}, \citenamefont {Hasegawa},\ and\
  \citenamefont {Aono}}]{tsuruoka}%
  \BibitemOpen
  \bibfield  {author} {\bibinfo {author} {\bibfnamefont {T.}~\bibnamefont
  {Tsuruoka}}, \bibinfo {author} {\bibfnamefont {K.}~\bibnamefont {Terabe}},
  \bibinfo {author} {\bibfnamefont {T.}~\bibnamefont {Hasegawa}}, \ and\
  \bibinfo {author} {\bibfnamefont {M.}~\bibnamefont {Aono}},\ }\href@noop {}
  {\bibfield  {journal} {\bibinfo  {journal} {Nanotechnology}\ }\textbf
  {\bibinfo {volume} {21}},\ \bibinfo {pages} {425205} (\bibinfo {year}
  {2010})}\BibitemShut {NoStop}%
\bibitem [{\citenamefont {Gu}\ \emph {et~al.}(2010)\citenamefont {Gu},
  \citenamefont {Tada},\ and\ \citenamefont {Watanabe}}]{ta2o5_oxide}%
  \BibitemOpen
  \bibfield  {author} {\bibinfo {author} {\bibfnamefont {T.}~\bibnamefont
  {Gu}}, \bibinfo {author} {\bibfnamefont {T.}~\bibnamefont {Tada}}, \ and\
  \bibinfo {author} {\bibfnamefont {S.}~\bibnamefont {Watanabe}},\ }\href@noop
  {} {\bibfield  {journal} {\bibinfo  {journal} {ACS Nano}\ }\textbf {\bibinfo
  {volume} {4}},\ \bibinfo {pages} {6477} (\bibinfo {year} {2010})}\BibitemShut
  {NoStop}%
\bibitem [{\citenamefont {Xu}\ \emph {et~al.}(2014)\citenamefont {Xu},
  \citenamefont {Liu},\ and\ \citenamefont {Anantram}}]{xu_alumina}%
  \BibitemOpen
  \bibfield  {author} {\bibinfo {author} {\bibfnamefont {X.}~\bibnamefont
  {Xu}}, \bibinfo {author} {\bibfnamefont {J.}~\bibnamefont {Liu}}, \ and\
  \bibinfo {author} {\bibfnamefont {M.~P.}\ \bibnamefont {Anantram}},\
  }\href@noop {} {\bibfield  {journal} {\bibinfo  {journal} {Journal of Applied
  Physics}\ }\textbf {\bibinfo {volume} {116}},\ \bibinfo {pages} {163701}
  (\bibinfo {year} {2014})}\BibitemShut {NoStop}%
\bibitem [{\citenamefont {Belmonte}\ \emph {et~al.}(2013)\citenamefont
  {Belmonte}, \citenamefont {Kim}, \citenamefont {Chan}, \citenamefont
  {Heylen}, \citenamefont {Fantini}, \citenamefont {Houssa}, \citenamefont
  {Jurczak},\ and\ \citenamefont {Goux}}]{al2o31}%
  \BibitemOpen
  \bibfield  {author} {\bibinfo {author} {\bibfnamefont {A.}~\bibnamefont
  {Belmonte}}, \bibinfo {author} {\bibfnamefont {W.}~\bibnamefont {Kim}},
  \bibinfo {author} {\bibfnamefont {B.}~\bibnamefont {Chan}}, \bibinfo {author}
  {\bibfnamefont {N.}~\bibnamefont {Heylen}}, \bibinfo {author} {\bibfnamefont
  {A.}~\bibnamefont {Fantini}}, \bibinfo {author} {\bibfnamefont
  {M.}~\bibnamefont {Houssa}}, \bibinfo {author} {\bibfnamefont
  {M.}~\bibnamefont {Jurczak}}, \ and\ \bibinfo {author} {\bibfnamefont
  {L.}~\bibnamefont {Goux}},\ }in\ \href@noop {} {\emph {\bibinfo {booktitle}
  {2013 5$^{th}$ IEEE International Memory Workshop}}}\ (\bibinfo {year}
  {2013})\ pp.\ \bibinfo {pages} {26--29}\BibitemShut {NoStop}%
\bibitem [{\citenamefont {Belmonte}\ \emph {et~al.}(2015)\citenamefont
  {Belmonte}, \citenamefont {Celano}, \citenamefont {Degraeve}, \citenamefont
  {Fantini}, \citenamefont {Redolfi}, \citenamefont {Vandervorst},
  \citenamefont {Houssa}, \citenamefont {Jurczak},\ and\ \citenamefont
  {Goux}}]{al2o32}%
  \BibitemOpen
  \bibfield  {author} {\bibinfo {author} {\bibfnamefont {A.}~\bibnamefont
  {Belmonte}}, \bibinfo {author} {\bibfnamefont {U.}~\bibnamefont {Celano}},
  \bibinfo {author} {\bibfnamefont {R.}~\bibnamefont {Degraeve}}, \bibinfo
  {author} {\bibfnamefont {A.}~\bibnamefont {Fantini}}, \bibinfo {author}
  {\bibfnamefont {A.}~\bibnamefont {Redolfi}}, \bibinfo {author} {\bibfnamefont
  {W.}~\bibnamefont {Vandervorst}}, \bibinfo {author} {\bibfnamefont
  {M.}~\bibnamefont {Houssa}}, \bibinfo {author} {\bibfnamefont
  {M.}~\bibnamefont {Jurczak}}, \ and\ \bibinfo {author} {\bibfnamefont
  {L.}~\bibnamefont {Goux}},\ }\href@noop {} {\bibfield  {journal} {\bibinfo
  {journal} {IEEE Electron Device Letters}\ }\textbf {\bibinfo {volume} {36}},\
  \bibinfo {pages} {775} (\bibinfo {year} {2015})}\BibitemShut {NoStop}%
\bibitem [{\citenamefont {Pandey}\ \emph {et~al.}(2015)\citenamefont {Pandey},
  \citenamefont {Meade},\ and\ \citenamefont {Sandhu}}]{sumeet}%
  \BibitemOpen
  \bibfield  {author} {\bibinfo {author} {\bibfnamefont {S.~C.}\ \bibnamefont
  {Pandey}}, \bibinfo {author} {\bibfnamefont {R.}~\bibnamefont {Meade}}, \
  and\ \bibinfo {author} {\bibfnamefont {G.~S.}\ \bibnamefont {Sandhu}},\
  }\href@noop {} {\bibfield  {journal} {\bibinfo  {journal} {Journal of Applied
  Physics}\ }\textbf {\bibinfo {volume} {117}},\ \bibinfo {pages} {054504}
  (\bibinfo {year} {2015})}\BibitemShut {NoStop}%
\bibitem [{\citenamefont {Tsai}\ \emph {et~al.}(2016)\citenamefont {Tsai},
  \citenamefont {Jiang}, \citenamefont {Ho}, \citenamefont {Lin},\ and\
  \citenamefont {Tseng}}]{bilayer_tsai}%
  \BibitemOpen
  \bibfield  {author} {\bibinfo {author} {\bibfnamefont {T.}~\bibnamefont
  {Tsai}}, \bibinfo {author} {\bibfnamefont {F.}~\bibnamefont {Jiang}},
  \bibinfo {author} {\bibfnamefont {C.}~\bibnamefont {Ho}}, \bibinfo {author}
  {\bibfnamefont {C.}~\bibnamefont {Lin}}, \ and\ \bibinfo {author}
  {\bibfnamefont {T.}~\bibnamefont {Tseng}},\ }\href@noop {} {\bibfield
  {journal} {\bibinfo  {journal} {IEEE Electron Device Letters}\ }\textbf
  {\bibinfo {volume} {37}},\ \bibinfo {pages} {1284} (\bibinfo {year}
  {2016})}\BibitemShut {NoStop}%
\bibitem [{\citenamefont {Barci}\ \emph {et~al.}(2016)\citenamefont {Barci},
  \citenamefont {Molas}, \citenamefont {Cagli}, \citenamefont {Vianello},
  \citenamefont {Bernard}, \citenamefont {Roule}, \citenamefont {Toffoli},
  \citenamefont {Cluzel}, \citenamefont {Salvo},\ and\ \citenamefont
  {Perniola}}]{bilayer_barci}%
  \BibitemOpen
  \bibfield  {author} {\bibinfo {author} {\bibfnamefont {M.}~\bibnamefont
  {Barci}}, \bibinfo {author} {\bibfnamefont {G.}~\bibnamefont {Molas}},
  \bibinfo {author} {\bibfnamefont {C.}~\bibnamefont {Cagli}}, \bibinfo
  {author} {\bibfnamefont {E.}~\bibnamefont {Vianello}}, \bibinfo {author}
  {\bibfnamefont {M.}~\bibnamefont {Bernard}}, \bibinfo {author} {\bibfnamefont
  {A.}~\bibnamefont {Roule}}, \bibinfo {author} {\bibfnamefont
  {A.}~\bibnamefont {Toffoli}}, \bibinfo {author} {\bibfnamefont
  {J.}~\bibnamefont {Cluzel}}, \bibinfo {author} {\bibfnamefont {B.~D.}\
  \bibnamefont {Salvo}}, \ and\ \bibinfo {author} {\bibfnamefont
  {L.}~\bibnamefont {Perniola}},\ }\href@noop {} {\bibfield  {journal}
  {\bibinfo  {journal} {IEEE Journal of the Electron Devices Society}\ }\textbf
  {\bibinfo {volume} {4}},\ \bibinfo {pages} {314} (\bibinfo {year}
  {2016})}\BibitemShut {NoStop}%
\bibitem [{\citenamefont {Kittl}\ \emph {et~al.}(2009)\citenamefont {Kittl},
  \citenamefont {Opsomer}, \citenamefont {Popovici}, \citenamefont {Menou},
  \citenamefont {Kaczer}, \citenamefont {Wang}, \citenamefont {Adelmann},
  \citenamefont {Pawlak}, \citenamefont {Tomida}, \citenamefont {Rothschild},\
  and\ \citenamefont {et~al.}}]{kittl_high_k}%
  \BibitemOpen
  \bibfield  {author} {\bibinfo {author} {\bibfnamefont {J.}~\bibnamefont
  {Kittl}}, \bibinfo {author} {\bibfnamefont {K.}~\bibnamefont {Opsomer}},
  \bibinfo {author} {\bibfnamefont {M.}~\bibnamefont {Popovici}}, \bibinfo
  {author} {\bibfnamefont {N.}~\bibnamefont {Menou}}, \bibinfo {author}
  {\bibfnamefont {B.}~\bibnamefont {Kaczer}}, \bibinfo {author} {\bibfnamefont
  {X.}~\bibnamefont {Wang}}, \bibinfo {author} {\bibfnamefont {C.}~\bibnamefont
  {Adelmann}}, \bibinfo {author} {\bibfnamefont {M.}~\bibnamefont {Pawlak}},
  \bibinfo {author} {\bibfnamefont {K.}~\bibnamefont {Tomida}}, \bibinfo
  {author} {\bibfnamefont {A.}~\bibnamefont {Rothschild}}, \ and\ \bibinfo
  {author} {\bibnamefont {et~al.}},\ }\href@noop {} {\bibfield  {journal}
  {\bibinfo  {journal} {Microelectronic Engineering}\ }\textbf {\bibinfo
  {volume} {86}},\ \bibinfo {pages} {1789 } (\bibinfo {year}
  {2009})}\BibitemShut {NoStop}%
\bibitem [{\citenamefont {Eklund}\ \emph {et~al.}(2009)\citenamefont {Eklund},
  \citenamefont {Sridharan}, \citenamefont {Singh},\ and\ \citenamefont
  {Bøttiger}}]{eklund_thermalstability_al2o3}%
  \BibitemOpen
  \bibfield  {author} {\bibinfo {author} {\bibfnamefont {P.}~\bibnamefont
  {Eklund}}, \bibinfo {author} {\bibfnamefont {M.}~\bibnamefont {Sridharan}},
  \bibinfo {author} {\bibfnamefont {G.}~\bibnamefont {Singh}}, \ and\ \bibinfo
  {author} {\bibfnamefont {J.}~\bibnamefont {Bøttiger}},\ }\href@noop {}
  {\bibfield  {journal} {\bibinfo  {journal} {Plasma Processes and Polymers}\
  }\textbf {\bibinfo {volume} {6}},\ \bibinfo {pages} {S907} (\bibinfo {year}
  {2009})}\BibitemShut {NoStop}%
\bibitem [{\citenamefont {Prasai}\ \emph {et~al.}(2018)\citenamefont {Prasai},
  \citenamefont {Subedi}, \citenamefont {Ferris}, \citenamefont {Biswas},\ and\
  \citenamefont {Drabold}}]{rrl_prasai18}%
  \BibitemOpen
  \bibfield  {author} {\bibinfo {author} {\bibfnamefont {K.}~\bibnamefont
  {Prasai}}, \bibinfo {author} {\bibfnamefont {K.~N.}\ \bibnamefont {Subedi}},
  \bibinfo {author} {\bibfnamefont {K.}~\bibnamefont {Ferris}}, \bibinfo
  {author} {\bibfnamefont {P.}~\bibnamefont {Biswas}}, \ and\ \bibinfo {author}
  {\bibfnamefont {D.~A.}\ \bibnamefont {Drabold}},\ }\href@noop {} {\bibfield
  {journal} {\bibinfo  {journal} {physica status solidi (RRL) – Rapid
  Research Letters}\ }\textbf {\bibinfo {volume} {12}},\ \bibinfo {pages}
  {1800238} (\bibinfo {year} {2018})}\BibitemShut {NoStop}%
\bibitem [{\citenamefont {{Guti{\'e}rrez}}\ \emph {et~al.}(2000)\citenamefont
  {{Guti{\'e}rrez}}, \citenamefont {{Belonoshko}}, \citenamefont {{Ahuja}},\
  and\ \citenamefont {{Johansson}}}]{gutierrez}%
  \BibitemOpen
  \bibfield  {author} {\bibinfo {author} {\bibfnamefont {G.}~\bibnamefont
  {{Guti{\'e}rrez}}}, \bibinfo {author} {\bibfnamefont {A.~B.}\ \bibnamefont
  {{Belonoshko}}}, \bibinfo {author} {\bibfnamefont {R.}~\bibnamefont
  {{Ahuja}}}, \ and\ \bibinfo {author} {\bibfnamefont {B.}~\bibnamefont
  {{Johansson}}},\ }\href@noop {} {\bibfield  {journal} {\bibinfo  {journal}
  {Physical Review E}\ }\textbf {\bibinfo {volume} {61}},\ \bibinfo {pages}
  {2723} (\bibinfo {year} {2000})}\BibitemShut {NoStop}%
\bibitem [{\citenamefont {Vashishta}\ \emph {et~al.}(2008)\citenamefont
  {Vashishta}, \citenamefont {Kalia}, \citenamefont {Nakano},\ and\
  \citenamefont {Rino}}]{vasistha_al2o3}%
  \BibitemOpen
  \bibfield  {author} {\bibinfo {author} {\bibfnamefont {P.}~\bibnamefont
  {Vashishta}}, \bibinfo {author} {\bibfnamefont {R.~K.}\ \bibnamefont
  {Kalia}}, \bibinfo {author} {\bibfnamefont {A.}~\bibnamefont {Nakano}}, \
  and\ \bibinfo {author} {\bibfnamefont {J.~P.}\ \bibnamefont {Rino}},\
  }\href@noop {} {\bibfield  {journal} {\bibinfo  {journal} {Journal of Applied
  Physics}\ }\textbf {\bibinfo {volume} {103}},\ \bibinfo {pages} {083504}
  (\bibinfo {year} {2008})}\BibitemShut {NoStop}%
\bibitem [{\citenamefont {Miranda Hernández~J.G.}\ and\ \citenamefont
  {Rocha-Rangel}(2006)}]{cudoped_reference}%
  \BibitemOpen
  \bibfield  {author} {\bibinfo {author} {\bibfnamefont {A.}~\bibnamefont
  {Miranda Hernández~J.G.}, \bibfnamefont {Soto~Guzm\'an}}\ and\ \bibinfo
  {author} {\bibfnamefont {E.}~\bibnamefont {Rocha-Rangel}},\ }\href@noop {}
  {\bibfield  {journal} {\bibinfo  {journal} {J.Ceram. Proc. Res}\ }\textbf
  {\bibinfo {volume} {7}},\ \bibinfo {pages} {311} (\bibinfo {year}
  {2006})}\BibitemShut {NoStop}%
\bibitem [{\citenamefont {Kresse}\ and\ \citenamefont {Hafner}(1993)}]{vasp}%
  \BibitemOpen
  \bibfield  {author} {\bibinfo {author} {\bibfnamefont {G.}~\bibnamefont
  {Kresse}}\ and\ \bibinfo {author} {\bibfnamefont {J.}~\bibnamefont
  {Hafner}},\ }\href@noop {} {\bibfield  {journal} {\bibinfo  {journal} {Phys.
  Rev. B}\ }\textbf {\bibinfo {volume} {47}},\ \bibinfo {pages} {558} (\bibinfo
  {year} {1993})}\BibitemShut {NoStop}%
\bibitem [{\citenamefont {Bl\"ochl}(1994)}]{PAW1}%
  \BibitemOpen
  \bibfield  {author} {\bibinfo {author} {\bibfnamefont {P.~E.}\ \bibnamefont
  {Bl\"ochl}},\ }\href@noop {} {\bibfield  {journal} {\bibinfo  {journal}
  {Phys. Rev. B}\ }\textbf {\bibinfo {volume} {50}},\ \bibinfo {pages} {17953}
  (\bibinfo {year} {1994})}\BibitemShut {NoStop}%
\bibitem [{\citenamefont {Kresse}\ and\ \citenamefont {Joubert}(1999)}]{PAW2}%
  \BibitemOpen
  \bibfield  {author} {\bibinfo {author} {\bibfnamefont {G.}~\bibnamefont
  {Kresse}}\ and\ \bibinfo {author} {\bibfnamefont {D.}~\bibnamefont
  {Joubert}},\ }\href@noop {} {\bibfield  {journal} {\bibinfo  {journal} {Phys.
  Rev. B}\ }\textbf {\bibinfo {volume} {59}},\ \bibinfo {pages} {1758}
  (\bibinfo {year} {1999})}\BibitemShut {NoStop}%
\bibitem [{\citenamefont {Perdew}\ and\ \citenamefont {Zunger}(1981)}]{LDA}%
  \BibitemOpen
  \bibfield  {author} {\bibinfo {author} {\bibfnamefont {J.~P.}\ \bibnamefont
  {Perdew}}\ and\ \bibinfo {author} {\bibfnamefont {A.}~\bibnamefont
  {Zunger}},\ }\href@noop {} {\bibfield  {journal} {\bibinfo  {journal} {Phys.
  Rev. B}\ }\textbf {\bibinfo {volume} {23}},\ \bibinfo {pages} {5048}
  (\bibinfo {year} {1981})}\BibitemShut {NoStop}%
\bibitem [{\citenamefont {Drabold}(2009)}]{dadepj}%
  \BibitemOpen
  \bibfield  {author} {\bibinfo {author} {\bibfnamefont {D.~A.}\ \bibnamefont
  {Drabold}},\ }\href@noop {} {\bibfield  {journal} {\bibinfo  {journal} {Eur.
  Phys. J. B}\ }\textbf {\bibinfo {volume} {68}},\ \bibinfo {pages} {1}
  (\bibinfo {year} {2009})}\BibitemShut {NoStop}%
\bibitem [{\citenamefont {Momida}\ \emph
  {et~al.}(2006{\natexlab{a}})\citenamefont {Momida}, \citenamefont {Hamada},
  \citenamefont {Takagi}, \citenamefont {Yamamoto}, \citenamefont {Uda},\ and\
  \citenamefont {Ohno}}]{cooling_rate_al2o3}%
  \BibitemOpen
  \bibfield  {author} {\bibinfo {author} {\bibfnamefont {H.}~\bibnamefont
  {Momida}}, \bibinfo {author} {\bibfnamefont {T.}~\bibnamefont {Hamada}},
  \bibinfo {author} {\bibfnamefont {Y.}~\bibnamefont {Takagi}}, \bibinfo
  {author} {\bibfnamefont {T.}~\bibnamefont {Yamamoto}}, \bibinfo {author}
  {\bibfnamefont {T.}~\bibnamefont {Uda}}, \ and\ \bibinfo {author}
  {\bibfnamefont {T.}~\bibnamefont {Ohno}},\ }\href@noop {} {\bibfield
  {journal} {\bibinfo  {journal} {Phys. Rev. B}\ }\textbf {\bibinfo {volume}
  {73}},\ \bibinfo {pages} {054108} (\bibinfo {year}
  {2006}{\natexlab{a}})}\BibitemShut {NoStop}%
\bibitem [{\citenamefont {Kubo}(1957)}]{kubo}%
  \BibitemOpen
  \bibfield  {author} {\bibinfo {author} {\bibfnamefont {R.}~\bibnamefont
  {Kubo}},\ }\href@noop {} {\bibfield  {journal} {\bibinfo  {journal} {J. Phys.
  Soc. Jpn.}\ }\textbf {\bibinfo {volume} {12}},\ \bibinfo {pages} {570}
  (\bibinfo {year} {1957})}\BibitemShut {NoStop}%
\bibitem [{\citenamefont {{Greenwood}}(1958)}]{greenwood}%
  \BibitemOpen
  \bibfield  {author} {\bibinfo {author} {\bibfnamefont {D.~A.}\ \bibnamefont
  {{Greenwood}}},\ }\href@noop {} {\bibfield  {journal} {\bibinfo  {journal}
  {Proceedings of the Physical Society}\ }\textbf {\bibinfo {volume} {71}},\
  \bibinfo {pages} {585} (\bibinfo {year} {1958})}\BibitemShut {NoStop}%
\bibitem [{\citenamefont {Feenstra}\ and\ \citenamefont {Widom}(2012)}]{widom}%
  \BibitemOpen
  \bibfield  {author} {\bibinfo {author} {\bibfnamefont {R.~M.}\ \bibnamefont
  {Feenstra}}\ and\ \bibinfo {author} {\bibfnamefont {M.}~\bibnamefont
  {Widom}},\ }\href@noop {} {} (\bibinfo {year} {2012}),\ \bibinfo {note}
  {\url{www.andrew.cmu.edu/user/feenstra/wavetrans}}\BibitemShut {NoStop}%
\bibitem [{\citenamefont {Lamparter}\ and\ \citenamefont
  {Kniep}(1997)}]{LAMPARTER1997405}%
  \BibitemOpen
  \bibfield  {author} {\bibinfo {author} {\bibfnamefont {P.}~\bibnamefont
  {Lamparter}}\ and\ \bibinfo {author} {\bibfnamefont {R.}~\bibnamefont
  {Kniep}},\ }\href@noop {} {\bibfield  {journal} {\bibinfo  {journal} {Physica
  B: Condensed Matter}\ }\textbf {\bibinfo {volume} {234-236}},\ \bibinfo
  {pages} {405 } (\bibinfo {year} {1997})}\BibitemShut {NoStop}%
\bibitem [{\citenamefont {Landron}\ \emph {et~al.}(2001)\citenamefont
  {Landron}, \citenamefont {Hennet}, \citenamefont {Jenkins}, \citenamefont
  {Greaves}, \citenamefont {Coutures},\ and\ \citenamefont
  {Soper}}]{landron_liquid_alumina}%
  \BibitemOpen
  \bibfield  {author} {\bibinfo {author} {\bibfnamefont {C.}~\bibnamefont
  {Landron}}, \bibinfo {author} {\bibfnamefont {L.}~\bibnamefont {Hennet}},
  \bibinfo {author} {\bibfnamefont {T.~E.}\ \bibnamefont {Jenkins}}, \bibinfo
  {author} {\bibfnamefont {G.~N.}\ \bibnamefont {Greaves}}, \bibinfo {author}
  {\bibfnamefont {J.~P.}\ \bibnamefont {Coutures}}, \ and\ \bibinfo {author}
  {\bibfnamefont {A.~K.}\ \bibnamefont {Soper}},\ }\href@noop {} {\bibfield
  {journal} {\bibinfo  {journal} {Phys. Rev. Lett.}\ }\textbf {\bibinfo
  {volume} {86}},\ \bibinfo {pages} {4839} (\bibinfo {year}
  {2001})}\BibitemShut {NoStop}%
\bibitem [{\citenamefont {Chagarov}\ and\ \citenamefont
  {Kummel}(2009)}]{kummel}%
  \BibitemOpen
  \bibfield  {author} {\bibinfo {author} {\bibfnamefont {E.~A.}\ \bibnamefont
  {Chagarov}}\ and\ \bibinfo {author} {\bibfnamefont {A.~C.}\ \bibnamefont
  {Kummel}},\ }\href@noop {} {\bibfield  {journal} {\bibinfo  {journal} {The
  Journal of Chemical Physics}\ }\textbf {\bibinfo {volume} {130}},\ \bibinfo
  {pages} {124717} (\bibinfo {year} {2009})}\BibitemShut {NoStop}%
\bibitem [{\citenamefont {Momida}\ \emph
  {et~al.}(2006{\natexlab{b}})\citenamefont {Momida}, \citenamefont {Hamada},
  \citenamefont {Takagi}, \citenamefont {Yamamoto}, \citenamefont {Uda},\ and\
  \citenamefont {Ohno}}]{momida}%
  \BibitemOpen
  \bibfield  {author} {\bibinfo {author} {\bibfnamefont {H.}~\bibnamefont
  {Momida}}, \bibinfo {author} {\bibfnamefont {T.}~\bibnamefont {Hamada}},
  \bibinfo {author} {\bibfnamefont {Y.}~\bibnamefont {Takagi}}, \bibinfo
  {author} {\bibfnamefont {T.}~\bibnamefont {Yamamoto}}, \bibinfo {author}
  {\bibfnamefont {T.}~\bibnamefont {Uda}}, \ and\ \bibinfo {author}
  {\bibfnamefont {T.}~\bibnamefont {Ohno}},\ }\href@noop {} {\bibfield
  {journal} {\bibinfo  {journal} {Phys. Rev. B}\ }\textbf {\bibinfo {volume}
  {73}},\ \bibinfo {pages} {054108} (\bibinfo {year}
  {2006}{\natexlab{b}})}\BibitemShut {NoStop}%
\bibitem [{\citenamefont {Sankaran}\ \emph {et~al.}(2012)\citenamefont
  {Sankaran}, \citenamefont {Goux}, \citenamefont {Clima}, \citenamefont
  {Mees}, \citenamefont {Kittl}, \citenamefont {Jurczak}, \citenamefont
  {Altimime}, \citenamefont {Rignanese},\ and\ \citenamefont
  {Pourtois}}]{sankaran}%
  \BibitemOpen
  \bibfield  {author} {\bibinfo {author} {\bibfnamefont {K.}~\bibnamefont
  {Sankaran}}, \bibinfo {author} {\bibfnamefont {L.}~\bibnamefont {Goux}},
  \bibinfo {author} {\bibfnamefont {S.}~\bibnamefont {Clima}}, \bibinfo
  {author} {\bibfnamefont {M.}~\bibnamefont {Mees}}, \bibinfo {author}
  {\bibfnamefont {J.}~\bibnamefont {Kittl}}, \bibinfo {author} {\bibfnamefont
  {M.}~\bibnamefont {Jurczak}}, \bibinfo {author} {\bibfnamefont
  {L.}~\bibnamefont {Altimime}}, \bibinfo {author} {\bibfnamefont {G.-M.}\
  \bibnamefont {Rignanese}}, \ and\ \bibinfo {author} {\bibfnamefont
  {G.}~\bibnamefont {Pourtois}}\ }(\bibinfo  {publisher} {Electrochemical
  Society},\ \bibinfo {year} {2012})\ pp.\ \bibinfo {pages}
  {317--330}\BibitemShut {NoStop}%
\bibitem [{\citenamefont {Dawson}\ and\ \citenamefont
  {Robertson}(2016)}]{dawson_cu_in_al2o3}%
  \BibitemOpen
  \bibfield  {author} {\bibinfo {author} {\bibfnamefont {J.~A.}\ \bibnamefont
  {Dawson}}\ and\ \bibinfo {author} {\bibfnamefont {J.}~\bibnamefont
  {Robertson}},\ }\href@noop {} {\bibfield  {journal} {\bibinfo  {journal} {The
  Journal of Physical Chemistry C}\ }\textbf {\bibinfo {volume} {120}},\
  \bibinfo {pages} {14474} (\bibinfo {year} {2016})}\BibitemShut {NoStop}%
\bibitem [{\citenamefont {Ziman}(1979)}]{ipr}%
  \BibitemOpen
  \bibfield  {author} {\bibinfo {author} {\bibfnamefont {J.~M.}\ \bibnamefont
  {Ziman}},\ }\href@noop {} {{\selectlanguage {English}\emph {\bibinfo {title}
  {Models of disorder : The theoretical physics of homogeneously disordered
  systems}}}}\ (\bibinfo  {publisher} {Cambridge University Press Cambridge
  [Eng.] ; New York},\ \bibinfo {year} {1979})\ pp.\ \bibinfo {pages} {xiii,
  525 p. :}\BibitemShut {NoStop}%
\bibitem [{\citenamefont {Prasai}\ \emph {et~al.}(2017)\citenamefont {Prasai},
  \citenamefont {Chen},\ and\ \citenamefont {Drabold}}]{prasai_gese3}%
  \BibitemOpen
  \bibfield  {author} {\bibinfo {author} {\bibfnamefont {K.}~\bibnamefont
  {Prasai}}, \bibinfo {author} {\bibfnamefont {G.}~\bibnamefont {Chen}}, \ and\
  \bibinfo {author} {\bibfnamefont {D.~A.}\ \bibnamefont {Drabold}},\
  }\href@noop {} {\bibfield  {journal} {\bibinfo  {journal} {Phys. Rev.
  Materials}\ }\textbf {\bibinfo {volume} {1}},\ \bibinfo {pages} {015603}
  (\bibinfo {year} {2017})}\BibitemShut {NoStop}%
\bibitem [{\citenamefont {Tang}\ \emph {et~al.}(2009)\citenamefont {Tang},
  \citenamefont {Sanville},\ and\ \citenamefont {Henkelman}}]{bader_code_ref}%
  \BibitemOpen
  \bibfield  {author} {\bibinfo {author} {\bibfnamefont {W.}~\bibnamefont
  {Tang}}, \bibinfo {author} {\bibfnamefont {E.}~\bibnamefont {Sanville}}, \
  and\ \bibinfo {author} {\bibfnamefont {G.}~\bibnamefont {Henkelman}},\
  }\href@noop {} {\bibfield  {journal} {\bibinfo  {journal} {Journal of
  Physics: Condensed Matter}\ }\textbf {\bibinfo {volume} {21}},\ \bibinfo
  {pages} {084204} (\bibinfo {year} {2009})}\BibitemShut {NoStop}%
\bibitem [{\citenamefont {Mott}\ and\ \citenamefont
  {Davis}(1979)}]{mott_davis}%
  \BibitemOpen
  \bibfield  {author} {\bibinfo {author} {\bibfnamefont {N.~F.}\ \bibnamefont
  {Mott}}\ and\ \bibinfo {author} {\bibfnamefont {E.~A.}\ \bibnamefont
  {Davis}},\ }\href@noop {} {{\selectlanguage {English}\emph {\bibinfo {title}
  {Electronic processes in non-crystalline materials / by N.F. Mott and E.A.
  Davis}}}},\ \bibinfo {edition} {2nd}\ ed.\ (\bibinfo  {publisher} {Clarendon
  Press ; Oxford University Press Oxford : New York},\ \bibinfo {year} {1979})\
  pp.\ \bibinfo {pages} {xiv, 590 p. :}\BibitemShut {NoStop}%
\bibitem [{\citenamefont {Landau}(1932)}]{landau}%
  \BibitemOpen
  \bibfield  {author} {\bibinfo {author} {\bibfnamefont {L.~D.}\ \bibnamefont
  {Landau}},\ }\href@noop {} {\bibfield  {journal} {\bibinfo  {journal} {Phys.
  Z. Sowjetunion}\ }\textbf {\bibinfo {volume} {2}},\ \bibinfo {pages} {46}
  (\bibinfo {year} {1932})}\BibitemShut {NoStop}%
\bibitem [{\citenamefont {{Zener}}(1932)}]{zener}%
  \BibitemOpen
  \bibfield  {author} {\bibinfo {author} {\bibfnamefont {C.}~\bibnamefont
  {{Zener}}},\ }\href@noop {} {\bibfield  {journal} {\bibinfo  {journal}
  {Proceedings of the Royal Society of London Series A}\ }\textbf {\bibinfo
  {volume} {137}},\ \bibinfo {pages} {696} (\bibinfo {year}
  {1932})}\BibitemShut {NoStop}%
\bibitem [{\citenamefont {Pandey}\ \emph {et~al.}(2014)\citenamefont {Pandey},
  \citenamefont {Cai}, \citenamefont {Podraza},\ and\ \citenamefont
  {Drabold}}]{anup_B-Si}%
  \BibitemOpen
  \bibfield  {author} {\bibinfo {author} {\bibfnamefont {A.}~\bibnamefont
  {Pandey}}, \bibinfo {author} {\bibfnamefont {B.}~\bibnamefont {Cai}},
  \bibinfo {author} {\bibfnamefont {N.}~\bibnamefont {Podraza}}, \ and\
  \bibinfo {author} {\bibfnamefont {D.~A.}\ \bibnamefont {Drabold}},\
  }\href@noop {} {\bibfield  {journal} {\bibinfo  {journal} {Phys. Rev.
  Applied}\ }\textbf {\bibinfo {volume} {2}},\ \bibinfo {pages} {054005}
  (\bibinfo {year} {2014})}\BibitemShut {NoStop}%
\bibitem [{\citenamefont {Kabir}\ \emph {et~al.}(2004)\citenamefont {Kabir},
  \citenamefont {Mookerjee},\ and\ \citenamefont
  {Bhattacharya}}]{kabir_cu_clusters}%
  \BibitemOpen
  \bibfield  {author} {\bibinfo {author} {\bibfnamefont {M.}~\bibnamefont
  {Kabir}}, \bibinfo {author} {\bibfnamefont {A.}~\bibnamefont {Mookerjee}}, \
  and\ \bibinfo {author} {\bibfnamefont {A.~K.}\ \bibnamefont {Bhattacharya}},\
  }\href@noop {} {\bibfield  {journal} {\bibinfo  {journal} {Phys. Rev. A}\
  }\textbf {\bibinfo {volume} {69}},\ \bibinfo {pages} {043203} (\bibinfo
  {year} {2004})}\BibitemShut {NoStop}%
\bibitem [{\citenamefont {Prasai}\ and\ \citenamefont
  {Drabold}(2011)}]{binaya_chalcogen}%
  \BibitemOpen
  \bibfield  {author} {\bibinfo {author} {\bibfnamefont {B.}~\bibnamefont
  {Prasai}}\ and\ \bibinfo {author} {\bibfnamefont {D.~A.}\ \bibnamefont
  {Drabold}},\ }\href@noop {} {\bibfield  {journal} {\bibinfo  {journal} {Phys.
  Rev. B}\ }\textbf {\bibinfo {volume} {83}},\ \bibinfo {pages} {094202}
  (\bibinfo {year} {2011})}\BibitemShut {NoStop}%
\bibitem [{\citenamefont {Pasquarello}\ \emph {et~al.}(1998)\citenamefont
  {Pasquarello}, \citenamefont {Sarnthein},\ and\ \citenamefont
  {Car}}]{pasquarello}%
  \BibitemOpen
  \bibfield  {author} {\bibinfo {author} {\bibfnamefont {A.}~\bibnamefont
  {Pasquarello}}, \bibinfo {author} {\bibfnamefont {J.}~\bibnamefont
  {Sarnthein}}, \ and\ \bibinfo {author} {\bibfnamefont {R.}~\bibnamefont
  {Car}},\ }\href@noop {} {\bibfield  {journal} {\bibinfo  {journal} {Phys.
  Rev. B}\ }\textbf {\bibinfo {volume} {57}},\ \bibinfo {pages} {14133}
  (\bibinfo {year} {1998})}\BibitemShut {NoStop}%
\bibitem [{\citenamefont {{Subedi}}\ \emph {et~al.}(2019)\citenamefont
  {{Subedi}}, \citenamefont {{Prasai}},\ and\ \citenamefont
  {{Drabold}}}]{ks_2019}%
  \BibitemOpen
  \bibfield  {author} {\bibinfo {author} {\bibfnamefont {K.~N.}\ \bibnamefont
  {{Subedi}}}, \bibinfo {author} {\bibfnamefont {K.}~\bibnamefont {{Prasai}}},
  \ and\ \bibinfo {author} {\bibfnamefont {D.~A.}\ \bibnamefont {{Drabold}}},\
  }\href@noop {} {\bibfield  {journal} {\bibinfo  {journal} {arXiv e-prints}\
  ,\ \bibinfo {eid} {arXiv:1901.04324}} (\bibinfo {year} {2019})},\ \Eprint
  {http://arxiv.org/abs/1901.04324} {arXiv:1901.04324 [cond-mat.dis-nn]}
  \BibitemShut {NoStop}%
\bibitem [{\citenamefont {Chaudhuri}\ \emph {et~al.}(2009)\citenamefont
  {Chaudhuri}, \citenamefont {Inam},\ and\ \citenamefont
  {Drabold}}]{choudary_Ag_dynamics}%
  \BibitemOpen
  \bibfield  {author} {\bibinfo {author} {\bibfnamefont {I.}~\bibnamefont
  {Chaudhuri}}, \bibinfo {author} {\bibfnamefont {F.}~\bibnamefont {Inam}}, \
  and\ \bibinfo {author} {\bibfnamefont {D.~A.}\ \bibnamefont {Drabold}},\
  }\href@noop {} {\bibfield  {journal} {\bibinfo  {journal} {Phys. Rev. B}\
  }\textbf {\bibinfo {volume} {79}},\ \bibinfo {pages} {100201(R)} (\bibinfo
  {year} {2009})}\BibitemShut {NoStop}%
\end{thebibliography}
%

\end{document}